\documentclass[pdflatex,sn-nature]{sn-jnl}

\usepackage{graphicx}%
\usepackage{multirow}%
\usepackage{amsmath,amssymb,amsfonts}%
\usepackage{amsthm}%
\usepackage{mathrsfs}%
\usepackage[title]{appendix}%
\usepackage{xcolor}%
\usepackage{textcomp}%
\usepackage{manyfoot}%
\usepackage{booktabs}%
\usepackage{longtable}%
\usepackage{pdflscape}%
\usepackage{algorithm}%
\usepackage{algorithmicx}%
\usepackage{algpseudocode}%
\usepackage{listings}%
\usepackage{placeins}%
\lstdefinestyle{mmcode}{%
  basicstyle=\footnotesize\ttfamily,%
  commentstyle=\footnotesize\ttfamily\itshape\color{gray!70!black},%
  showstringspaces=false,%
  columns=fullflexible,%
  keepspaces=true,%
  breaklines=true,%
  frame=single,%
  framesep=4pt,%
  framerule=0.4pt,%
  rulecolor=\color{gray!40},%
  xleftmargin=6pt,%
  xrightmargin=6pt,%
  aboveskip=4pt,%
  belowskip=4pt,%
}%
\hypersetup{hypertexnames=false}%

\theoremstyle{thmstyleone}%
\theoremstyle{thmstyletwo}%

\theoremstyle{thmstylethree}%

\begin{document}
\title[MultiMolecule]{MultiMolecule: a modular ecosystem for biomolecular sequence-model workflows}


\author*[1]{\fnm{Zhiyuan} \sur{Chen}}\email{this@zyc.ai}

\affil*[1]{\orgname{DanLing Team}}


\abstract{
  Biomolecular sequence models are increasingly reused outside the studies in which they were introduced, but public checkpoints rarely preserve the execution context needed to inspect source-defined behavior, adapt models to new assays, compare models under shared task definitions or deploy biological predictions.
  MultiMolecule is an open-source Python ecosystem that turns heterogeneous RNA, DNA and protein sequence-model releases into complete, source-checked model-family implementations with shared loading, workflow and prediction interfaces.
  The Resource state reported here includes 53 complete model-family implementations with 112 standardized model checkpoints, together with 16 curated dataset resources released through 39 public dataset repositories and 10 user-facing prediction pipelines.
  Standardized components are linked to source provenance, conversion or preparation code, source-reference checks, Extended Data summaries and public documentation, allowing users to inspect what was standardized, what behavior was checked and how each component enters training, evaluation, inference or deployment.
  By shifting reuse from repository-specific checkpoints to executable implementations connected to standardized checkpoints, curated datasets, Runner workflows and biological prediction pipelines, MultiMolecule provides common infrastructure for preserving source-defined model behavior, adapting models to new assays, enabling controlled evaluation and deploying biomolecular predictions.
}

\keywords{machine learning, biomolecular sequence models, RNA, DNA, protein}



\maketitle

\pagebreak

\section{Introduction}\label{sec:introduction}

Biomolecular sequence models have become working instruments for predicting molecular structure, regulation, processing, stability and function from biological sequences.
They are no longer merely outputs of individual studies, but starting points for assay adaptation, controlled evaluation and deployment across RNA, DNA and protein workflows.
For such reuse, public availability is necessary but not sufficient.
A checkpoint, and often even its source repository, rarely defines the execution context needed to make a model inspectable, comparable, adaptable and deployable outside its original release context.

Practical reuse is limited by how model behavior is packaged at release.
A released predictor commonly couples its learned parameters to a specific runtime environment, framework version, checkpoint-distribution channel and hardware-dependent execution path.
Reconstructing this context is often necessary before the model can be applied outside its original study.
Even after installation succeeds, users still need evidence that a standardized or reused implementation preserves source-defined behavior.
As a consequence, assay adaptation, controlled evaluation and deployment depend as much on recovering release-specific execution context as on obtaining the checkpoint.

Source-defined behavior is distributed across weights, code and surrounding conventions.
It is not contained in the weight tensor alone, but in the coupled sequence encoding, checkpoint layout, model implementation, task-specific heads, output semantics and evaluation procedure that surround it.
Silent changes in any of these components can alter downstream quantities while preserving the appearance of using the same published model.
For a field that increasingly compares model families and molecular modalities, implementation details become controlled scientific variables rather than merely engineering choices.
A reusable Resource must therefore preserve source-defined behavior, expose maintained execution interfaces and provide public evidence that standardization retains the quantities used for analysis.

Execution fragility compounds as software ecosystems age.
Many useful biomolecular models were released under the software assumptions of their original studies, including older deep-learning libraries, pinned dependency stacks, version-specific accelerator behavior, framework-specific kernels and installation procedures that are difficult to reproduce on current systems.
A model can remain scientifically valuable while becoming hard to install, run on modern hardware, check against source behavior or integrate with current machine-learning tools.
This is especially important in biomolecular sequence modeling, where small task-specific predictors, conventional neural architectures and newer pretrained backbones all remain useful for different biological questions.
The field therefore needs infrastructure that preserves older and newer models as executable scientific objects, not merely as archived checkpoints.

Existing infrastructure provides essential foundations but does not by itself solve this cross-family execution problem.
Tensor libraries and model hubs make computation and checkpoint sharing easier \cite{paszke2019pytorch,wolf2020transformers}.
Workflow systems improve provenance and portability for multi-step analyses \cite{wratten2021reproducible}.
Scalable biomolecular AI frameworks emphasize distributed training and high-throughput model development \cite{stjohn2024bionemo}.
Faithful single-architecture reimplementations can deeply reproduce one influential model family \cite{ahdritz2024openfold}.
Model-zoo resources established the value of shared access to trained predictors \cite{avsec2019kipoi}.
The unresolved gap is a maintained layer that connects biological inputs, source-linked implementations, standardized checkpoints, executable workflows, prediction pipelines and validation evidence across heterogeneous biomolecular model families.

MultiMolecule addresses this gap by treating reuse as documented executable infrastructure rather than as weight sharing alone.
It organizes biomolecular sequence-model reuse around complete model-family implementations, standardized checkpoints, curated dataset resources, Runner workflows, prediction pipelines and validation evidence.
The Resource treats task-specific predictors, conventional deep-learning models and pretrained sequence backbones as first-class model families under the same implementation, validation and deployment conventions.
Each complete model-family implementation preserves the source model-facing interface under shared MultiMolecule conventions, while associated checkpoints are converted into standardized formats with provenance and source-reference checks.
Curated dataset resources, biological sequence I/O, molecular tokenization, reusable neural modules, Runner workflows and user-facing prediction pipelines are organized around these implementations rather than around model-specific scripts.
This design makes model reuse something users can execute, inspect and extend, rather than a reconstruction task left to undocumented repository conventions.

Here we present MultiMolecule as an executable Resource for biomolecular sequence modeling and evaluate it at the level of reusable infrastructure.
The Resource state reported here includes 53 complete model-family implementations with 112 standardized model checkpoints, together with 16 curated dataset resources released through 39 public dataset repositories and 10 user-facing prediction pipelines.
We assess the Resource where users reuse its components---implementations, checkpoints, workflows, pipelines and validation evidence---rather than as a benchmark of biological model accuracy.
MultiMolecule provides shared infrastructure for preserving source-defined behavior, auditing standardization, adapting models to new assays, enabling controlled evaluation, extending model families and deploying biological predictions as reusable scientific infrastructure.

\section{Results}\label{sec:results}

\begin{figure}[!htbp]
  \centering
  \includegraphics[width=\linewidth]{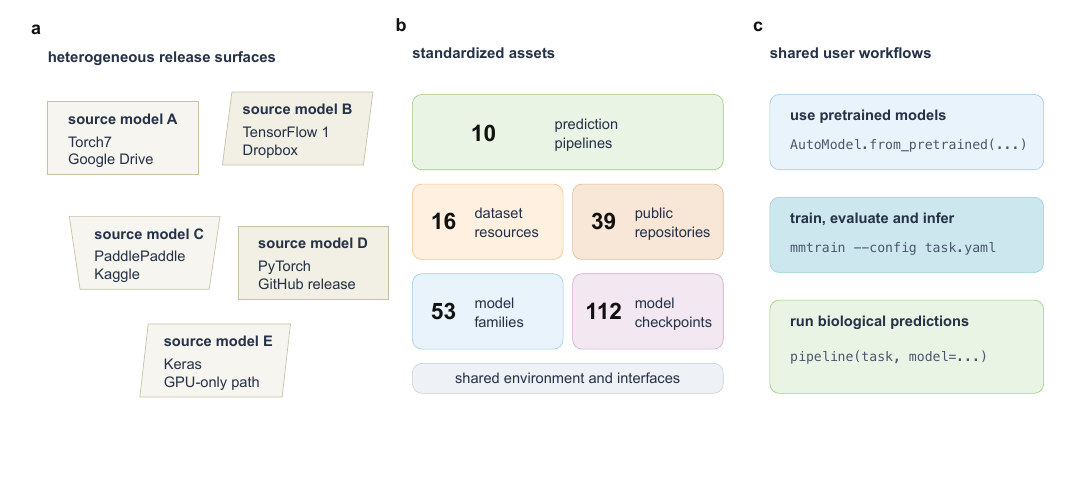}
  \caption{\textbf{MultiMolecule organizes biomolecular sequence-model reuse as an executable ecosystem.}
    \textbf{a,} Representative source-release archetypes illustrate the heterogeneous software, checkpoint-distribution, runtime and inference surfaces exposed by published model releases.
    \textbf{b,} MultiMolecule reorganizes these heterogeneous release surfaces into a connected, standardized ecosystem with a manuscript-reported inventory of 53 complete model-family implementations, 112 standardized model checkpoints, 16 curated dataset resources released through 39 public dataset repositories and 10 user-facing prediction pipelines.
    These assets enter shared loading, execution, validation and documentation conventions rather than remaining independent catalogue entries.
  \textbf{c,} Users enter the ecosystem through shared model-loading APIs, configuration-driven Runner workflows and task-oriented prediction pipelines, replacing separate per-model setup paths with reusable model, workflow and deployment interfaces.}
  \label{fig:ecosystem}
\end{figure}

\subsection{MultiMolecule organizes biomolecular sequence-model reuse as an executable ecosystem}\label{subsec:ecosystem}

MultiMolecule is released as a connected ecosystem of standardized model and dataset assets, workflow and prediction interfaces, with validation evidence summarized in Extended Data.
The Resource state reported here is organized into five operational layers: complete model-family implementations, standardized model checkpoints, curated dataset resources, Runner workflows and user-facing prediction pipelines.
The manuscript-reported inventory comprises 53 complete model-family implementations with 112 standardized model checkpoints, 16 curated dataset resources released through 39 public dataset repositories and 10 user-facing prediction pipelines (Fig.~\ref{fig:ecosystem}).
Each counted asset enters the inventory only after it is linked to the corresponding implementation, conversion or preparation procedure, relevant validation evidence, public documentation and a documented route for loading, execution or deployment.
This organization replaces source-specific execution surfaces with shared loading, workflow and deployment interfaces.

These layers form one operational path from biological data and source-model releases to executable workflows.
Biological sequences and curated datasets are standardized through sequence I/O and molecular tokenization.
Complete model-family implementations provide common configuration, tokenizer, model-output and task-head interfaces.
Reusable modules provide shared model components used by these implementations and workflows.
Runner workflows compose datasets, checkpoints, models, criteria and metrics into training, evaluation and inference executions.
Prediction pipelines expose selected standardized checkpoints as deployment-facing biological tools.
Extended Data summaries and public documentation link each public asset to its source, conversion or preparation procedure, tested behavior and public model card, dataset card or interface documentation.

Inclusion is driven by executable completeness rather than catalogue size.
A model family enters the Resource when its implementation, standardized checkpoints, workflow interfaces and validation checks satisfy the same criteria.
This keeps the counted assets inspectable while preserving breadth across RNA, transcript, genomic, regulatory, splicing and protein sequence-model workflows.
The following sections describe the main layers of this ecosystem: curated dataset resources, complete model-family implementations, standardized checkpoints and conversion summaries, Runner-based workflows, user-facing prediction pipelines and Resource-level validation evidence.

\begin{figure}[!htbp]
  \centering
  \includegraphics[width=\linewidth]{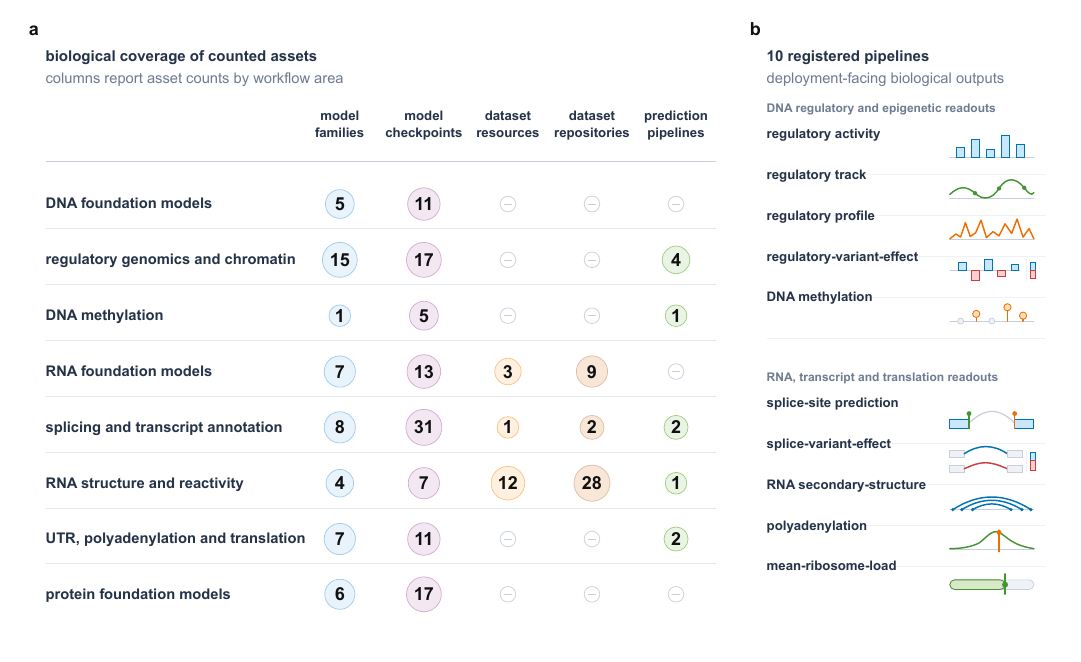}
  \caption{\textbf{MultiMolecule summarizes model, dataset and prediction coverage across biomolecular sequence workflows.}
    \textbf{a,} Coverage of the Resource state reported here across eight biological workflow areas: DNA foundation models; regulatory genomics and chromatin; DNA methylation; RNA foundation models; RNA structure and reactivity; splicing and transcript annotation; UTR, polyadenylation and translation; and protein foundation models.
    Rows correspond to these areas, and columns summarize counted assets: complete model-family implementations, standardized model checkpoints, curated dataset resources, public dataset repositories and registered prediction pipelines.
    The public dataset repository column reports the number of released repositories through which the curated dataset resources of each area are distributed, not a separate workflow layer.
    Numbers indicate the released assets associated with each area; empty entries indicate that no asset of that class is counted for that area in the manuscript-reported inventory, and a workflow area need not be populated across every layer.
    \textbf{b,} Registered prediction pipelines grouped by biological readout.
    The registry spans regulatory-activity, regulatory-track, regulatory-profile and regulatory variant effect prediction, DNA methylation prediction, splice-site and splice variant effect prediction, RNA secondary-structure prediction, polyadenylation prediction and mean-ribosome-load prediction.
  Glyphs indicate the returned output geometry, including sequence-level scores, positional tracks or profiles, variant effect scores, pairwise/contact outputs and scalar regression outputs.}
  \label{fig:biological-coverage}
\end{figure}

\subsection{Curated dataset resources provide documented biological inputs and task schemas}\label{subsec:dataset-resources}

Curated dataset resources provide the data-facing layer of the MultiMolecule ecosystem.
The manuscript-reported inventory includes 16 curated dataset resources distributed through 39 public dataset repositories, with coverage across RNA foundation-model inputs, RNA structure and reactivity, splicing and transcript annotation and related transcript workflows (Fig.~\ref{fig:biological-coverage}).
Each dataset resource is released with source provenance, preprocessing or preparation code, processed fields, sequence columns, label schemas, redistribution status and public dataset-card documentation.
Dataset summaries make workflow wiring explicit rather than treating processed data as anonymous benchmark files.

For each curated resource, MultiMolecule documents how biological sequences, structure annotations, positional labels, pairwise labels or scalar readouts are represented and how they map to sequence-level, token-level or contact-level tasks.
These summaries allow Runner workflows to construct task descriptors and output schemas from documented data fields, while still requiring user-declared overrides when dataset semantics are ambiguous.
The curated datasets are workflow resources rather than universal leakage-free benchmarks.
They provide documented biological inputs and reusable processed datasets for training, evaluation and examples, but a leakage-free comparison holds only where a split policy explicitly provides one.
This scope keeps the dataset layer useful for executable workflows without overclaiming benchmark validity across arbitrary downstream comparisons.
Full dataset-resource summaries are provided in Extended Data Table~\ref{tab:ed-datasets}.

\subsection{Complete model-family implementations standardize heterogeneous source releases}\label{subsec:complete-implementations}

The complete model-family implementation is the model-level unit counted by MultiMolecule.
This boundary defines what must be standardized before a source release can be loaded, inspected, trained, evaluated or deployed through shared interfaces.
For each family, MultiMolecule preserves the model-facing surface used by downstream users, including configuration, tokenizer, model classes, output schema and source-exposed task heads.
Source-specific choices are handled inside the implementation and checkpoint-conversion layer rather than left as runtime assumptions for the user.

The completeness criterion excludes repository code that does not define the model API.
Repository-specific training harnesses, dataset loaders, visualization utilities, logging code and analysis scripts are outside this criterion unless they define the model-facing interface.
Each implementation registers standardized configuration, tokenizer, model-output and automatic model-loading interfaces under MultiMolecule conventions.
It also provides conversion code for redistributable checkpoint variants included in the Resource and includes conversion checks against source-reference outputs for the quantities exposed by that family.
All complete implementations expose the standardized model interface used by Runner workflows, whereas families with registered biological prediction tasks are additionally checked for prediction-pipeline compatibility.

The implementation boundary separates source fidelity from downstream use.
Source-specific choices, including input encoding, tensor naming, special-token handling, tied-weight conventions and task-head layouts, are handled during implementation and checkpoint conversion rather than exposed to users at runtime.
Downstream workflows therefore interact with stable MultiMolecule objects: a configuration, tokenizer, model class, output schema and task head that follow common conventions across families.
This separation allows users to switch among compatible model families without rewriting model-loading code, output-parsing logic or task-head construction.

Under this criterion, the Resource state reported here contains 53 complete model-family implementations spanning RNA, transcript, genomic, regulatory, splicing and protein sequence-model workflows, with biological coverage summarized in Fig.~\ref{fig:biological-coverage}.
The included families cover task-specific predictors, conventional deep-learning architectures and pretrained sequence backbones rather than privileging a single modeling paradigm.
They expose sequence-level, token-level, pairwise or contact-level, masked-language-modeling, causal-language-modeling and specialized prediction interfaces.
The full per-family completeness matrix is provided in Extended Data Table~\ref{tab:ed-completeness}.
Across this breadth, each counted family meets the same implementation-completeness criterion.

\begin{figure}[!htbp]
  \centering
  \includegraphics[width=\linewidth]{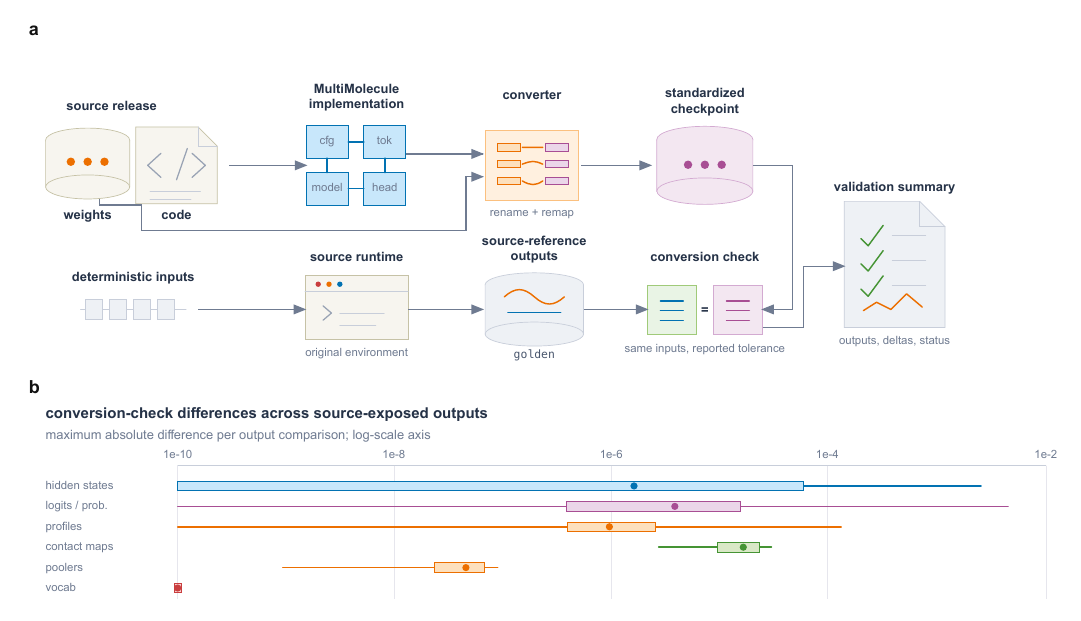}
  \caption{\textbf{Standardized model checkpoints are produced by reimplementation, conversion and source-reference checking.}
    \textbf{a,} Model standardization separates two coupled paths from each source release.
    Source model-facing code is reimplemented under MultiMolecule conventions to produce a complete model-family implementation, while source checkpoints are converted by family-specific scripts into standardized model checkpoints.
    In parallel, the original source code and source checkpoints are executed in the upstream environment on deterministic reference inputs to generate source-reference outputs, referred to as golden outputs in the committed source-reference fixtures.
    The standardized implementation and standardized checkpoint are then evaluated on the same inputs, and their outputs are compared with the corresponding source-reference outputs.
    \textbf{b,} Distribution of maximum absolute differences across 177 output-level conversion checks for the 112 standardized model checkpoints.
    Each check compares one source-exposed output field from a standardized checkpoint with the corresponding source-reference output under the reported numerical criterion.
  These summaries document what was converted, what source-defined behavior was checked and how closely standardized outputs agree with source-reference outputs.}
  \label{fig:checkpoint-conversion}
\end{figure}

The 53 implementations are accompanied by 112 standardized model checkpoints (Fig.~\ref{fig:checkpoint-conversion}).
The checkpoint-conversion summaries contain 177 output-level source-reference checks across these checkpoints, spanning 21 source-exposed output fields.
Depending on the source family, the compared fields include hidden states, logits, contact maps, pooler outputs, vocabulary embeddings, coverage or profile tracks and task-head predictions.
Each check evaluates source-reference outputs and the corresponding standardized MultiMolecule outputs on deterministic reference inputs and reports element-wise absolute-difference summaries, including the minimum, 25th percentile, median, 75th percentile, maximum and mean.
All 177 output-level checks are within their reported conversion criteria.
When exact equality is not expected because of floating-point precision, hardware differences or dependency-version effects, the tolerance and comparison metric are reported rather than treated as implicit implementation details.
Output-level conversion summaries are provided in Extended Data Table~\ref{tab:ed-checkpoints}.

Input-encoding differences are resolved during checkpoint conversion.
When a source release uses family-specific tokenization, special-token placement, codon representations or tokenizer-side conventions, MultiMolecule encodes the upstream-to-standardized mapping during conversion and distributes the converted tokenizer files with the standardized checkpoint.
Runtime users interact with the standardized MultiMolecule tokenizer and model interface, while conversion summaries document agreement with source-reference behavior.
Token-ID equality with the source tokenizer is not required when a deliberate vocabulary remapping has been applied; output agreement after conversion is the relevant validation target.

Complete implementations provide the model boundary on which the rest of the Resource depends.
Runner workflows rely on implementation-level consistency to execute standardized model families under explicit task definitions.
Prediction pipelines rely on implementation-level consistency to resolve input processing, load compatible checkpoints and format structured biological outputs.
Model cards, conversion scripts and Extended Data summaries link each standardized checkpoint to its source checkpoint and source-reference checks.
Thus, complete model-family implementations provide the model-facing substrate for MultiMolecule's executable ecosystem.

\subsection{Runner workflows execute standardized implementations under explicit task definitions}\label{subsec:runner-workflows}

\begin{figure}[!htbp]
  \centering
  \includegraphics[width=\linewidth]{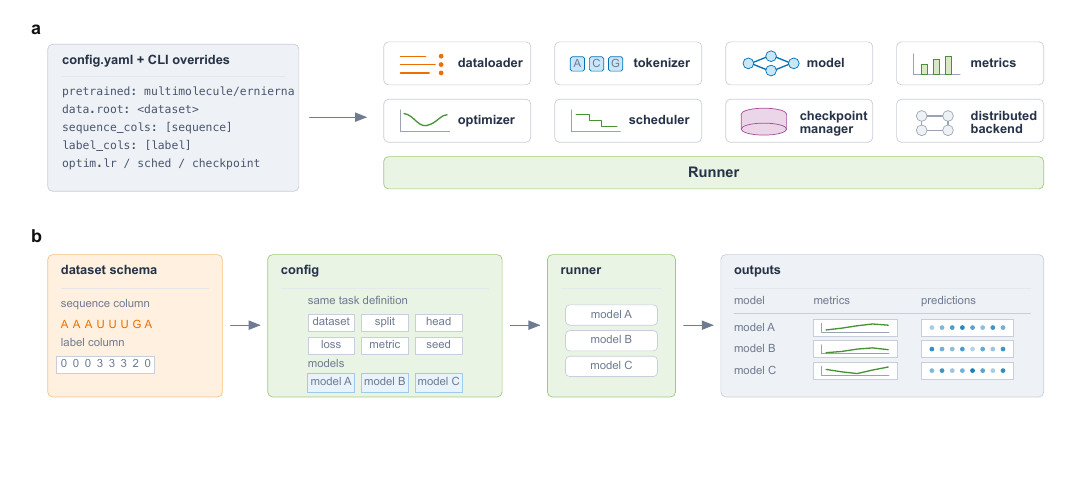}
  \caption{\textbf{Runner workflows execute standardized model implementations under explicit task definitions.}
    \textbf{a,} A Runner workflow starts from a declarative recipe that specifies the data source, sequence and label schema, standardized checkpoint, task setting and execution parameters.
    From this recipe, MultiMolecule resolves input processing, the standardized model implementation, task-specific execution components and output schemas required for training, evaluation or inference, and writes the resulting workflow artifacts.
    \textbf{b,} Controlled model evaluation keeps the dataset and task definition fixed while the model family or standardized checkpoint is varied.
  The Runner executes each model under the same configured workflow and writes the resulting metrics and prediction outputs, so varying the model family within one task definition preserves the same data-processing and evaluation path.}
  \label{fig:runner-workflows}
\end{figure}

Runner workflows turn standardized implementations and checkpoints into executable dataset-level analyses.
MultiMolecule provides this execution layer through Runner workflows (Fig.~\ref{fig:runner-workflows}) that bind biological inputs, standardized model implementations, task definitions, evaluation settings and output schemas into executable training, evaluation and inference workflows.
The Runner makes standardized implementations and checkpoints executable on biological datasets under shared data, task and evaluation conventions, rather than leaving each family tied to its original scripts.

A Runner workflow starts from a declarative configuration rather than a model-specific training script.
The configuration specifies the data source, split structure, sequence and label schema, standardized checkpoint or newly initialized backbone, optional modules, optimization settings, random seed, checkpoint restoration and execution mode.
From this configuration, the Runner resolves input processing, constructs datasets and data loaders, instantiates the standardized model implementation, assembles task-specific execution components and executes the requested workflow.
Because these choices are specified explicitly, the same workflow can be rerun, inspected, modified and shared without recovering hidden assumptions from a source repository.

The defining property of the Runner is explicit task-defined execution.
The Runner separates standardized model-family reuse from the task-specific source scripts that originally accompanied each family.
A dataset task definition specifies the prediction level, statistical objective, label dimensionality, metrics and output schema used for training, evaluation or inference.
The same task definition can then be executed with different standardized checkpoints or model families through identical data processing and evaluation logic.
Changing the task definition creates a separate workflow, whereas changing the model family within one task definition preserves the configured workflow.
This design makes model-task pairing an explicit experimental choice: MultiMolecule standardizes how the experiment is run, while the configured assay labels and metrics determine whether a model is biologically useful for that task.
Any downstream comparison therefore reflects the model implementation and standardized checkpoint, not hidden differences among preprocessing scripts, task setup or evaluation harnesses.

Internal task descriptors provide the handoff from data interpretation to model assembly.
Within the Runner, dataset labels are resolved into typed descriptors that specify output level, statistical objective, label dimensionality and missing-label behavior.
These descriptors determine compatible heads, losses, metrics and output schemas, allowing datasets with sequence-level labels, per-position labels or pairwise labels to be mapped to the appropriate model components.
When the schema is ambiguous, user-declared overrides are required rather than silently assigning biological meaning from geometry alone.
This keeps task semantics attached to the workflow while avoiding model-specific preprocessing code.

Reuse evidence is separated by Resource layer.
Downstream RNA modeling studies, including BEACON, mRNABERT and mRNABench, reuse MultiMolecule implementations, checkpoints or compatible model families under independent benchmark or training stacks \cite{ren2024beacon,xiong2025mrnabert,shi2026mrnabench}.
These studies are cited as implementation-layer reuse evidence rather than as Runner-validation evidence.
Runner-level evidence comes from full-stack workflows that execute through the Runner itself.
CHANRG provides a full-stack RNA secondary-structure workflow built on MultiMolecule Runner workflows and executed under a shared structured-prediction setting \cite{chen2026chanrg}.
Uni-RNA secondary-structure training provides a second full-stack structured-prediction workflow using the MultiMolecule data, model, Runner and evaluation stack \cite{wang2023unirna}.
The repository's in-tree examples additionally exercise the data-loading, model-loading, pipeline and Runner entry points, including a contact-level Runner workflow.
For each representative Runner workflow, the manuscript reports the resolved configuration, input processor, standardized checkpoint, internal task descriptor, task-specific components, metrics, output schema, produced files and software environment.
The resulting artifacts include resolved configurations, checkpoints, metrics files, prediction files and exported model directories.
Each workflow therefore yields inspectable execution outputs linking data, model, task definition and outputs.

By making workflows configuration-driven, MultiMolecule separates model reuse from source-repository reconstruction.
Users can fine-tune standardized checkpoints, evaluate compatible model families under shared definitions or export trained checkpoints for deployment without rewriting input handling, task-specific execution logic or output formatting.
Developers can add new complete model-family implementations or reusable modules while retaining the same execution layer.
The next section describes how selected Runner-compatible and pretrained checkpoints are exposed through user-facing prediction pipelines for direct biological use.

\subsection{Prediction pipelines expose standardized checkpoints as biological tools}\label{subsec:pipelines}

\begin{figure}[!htbp]
  \centering
  \includegraphics[width=\linewidth]{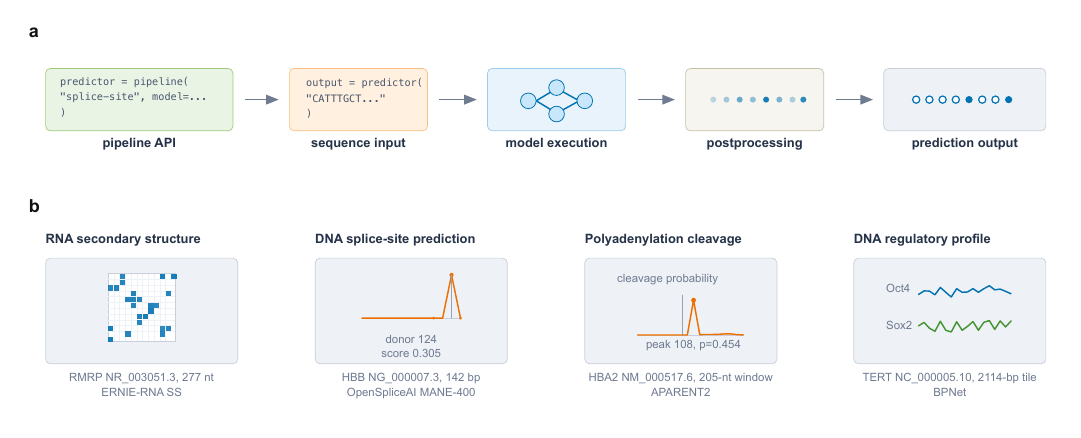}
  \caption{\textbf{Prediction pipelines expose standardized checkpoints as directly usable biological prediction tools.}
    \textbf{a,} A MultiMolecule prediction pipeline is invoked through the Transformers \texttt{pipeline} interface using a registered task string and a compatible model identifier.
    The resulting callable accepts sequence input, executes the standardized model implementation and task-specific postprocessing, and returns structured prediction output.
    The registered prediction surfaces are summarized in Fig.~\ref{fig:biological-coverage}; here the focus is how a user invokes the pipeline interface and what form of biological output is returned.
    \textbf{b,} Representative pipeline examples show four outputs across RNA and DNA inputs: RNA secondary-structure prediction, DNA splice-site prediction, polyadenylation cleavage-site prediction and DNA regulatory-profile prediction.
    The polyadenylation example displays a cleavage-position probability distribution, and the regulatory-profile example displays position-resolved regulatory signal tracks following the plotting conventions used by the public apps.
  The example values are actual outputs produced by the released pipeline interfaces on named biological inputs, demonstrating the returned biological prediction format rather than comparative task performance.}
  \label{fig:pipelines}
\end{figure}

Prediction pipelines turn standardized checkpoints into direct biological prediction interfaces.
Standardized model implementations and Runner workflows address model reuse for training, evaluation and inference, but many users need a simpler deployment surface for direct biological prediction.
MultiMolecule provides this surface through user-facing prediction pipelines (Fig.~\ref{fig:pipelines}) that bind a biological task name to accepted inputs, compatible standardized checkpoints, tokenizer preparation, model execution, postprocessing and structured outputs.
The pipeline layer is the deployment-facing counterpart of the Runner: it exposes standardized checkpoints through biological task interfaces rather than through model-internal tensor conventions.
This design allows pretrained or Runner-fine-tuned checkpoints to be used for prediction without reconstructing model-specific inference scripts.

The manuscript-reported inventory provides user-facing prediction pipelines across RNA structure, splicing, polyadenylation, regulatory genomics, variant effects, methylation and translation-related prediction.
These pipelines include RNA secondary-structure prediction, splice-site prediction, splice variant effect prediction, polyadenylation prediction, regulatory-activity prediction, regulatory-track prediction, regulatory-profile prediction, regulatory variant effect prediction, DNA methylation prediction and mean-ribosome-load prediction.
Each registered pipeline declares accepted biological inputs, compatible standardized checkpoints, postprocessing rules, returned biological outputs and the public interface through which the task can be used.
The 10 registered pipelines are currently exposed through nine public Hugging Face Spaces because the regulatory-track and regulatory-profile pipelines share a single regulatory-signal demo surface.
This registry separates the biological task interface from model-family implementation details.
Different compatible standardized checkpoints can enter the same user-facing task when they expose the required standardized output interface, and a complete model-family implementation can support multiple prediction surfaces when it includes the corresponding heads and postprocessing rules.
The complete prediction-pipeline registry is provided in Extended Data Table~\ref{tab:ed-pipelines}.

Representative released pipelines were executed on named RNA and DNA inputs to show that the registered interfaces return structured biological predictions rather than only registry entries (Fig.~\ref{fig:pipelines}).
The RNA secondary-structure pipeline illustrates how a standardized model output becomes a biological result.
A compatible standardized checkpoint receives an RNA sequence, applies the registered tokenizer, executes the standardized pairwise prediction head and returns contact probabilities or a contact map over biological positions.
The same pairwise output can be postprocessed into dot-bracket notation when a compact structure representation is useful.
The pipeline hides model-internal details such as special tokens, padding positions, tensor layouts and head names while preserving model-derived quantities for downstream analysis.
This worked example shows how a registered pipeline turns a standardized pairwise output into a deployable biological structure prediction.

The other registered pipelines follow the same pattern.
Splice pipelines map sequence or variant inputs to splice-site or splice variant effect outputs.
Regulatory pipelines expose sequence-level, track-level, profile-level and variant effect predictions under separate output schemas.
Methylation, polyadenylation and mean-ribosome-load pipelines map compatible sequence inputs to task-specific probability or regression outputs.
The biological tasks differ, but the user-facing pattern is shared: a user selects a registered task and a compatible standardized checkpoint, and the pipeline resolves the tokenizer, model implementation, output schema and postprocessing logic.
This shared interface prevents each task family from requiring a separate inference implementation.

Each registered pipeline is documented by its registry entry and implementation tests.
For each registered prediction interface, pipeline tests check accepted input forms, compatible standardized implementations and checkpoints, output-field structure and deterministic postprocessing.
These checks complement the checkpoint-conversion checks described above: conversion summaries document agreement with selected source-exposed model behavior, whereas pipeline tests establish that registered biological prediction interfaces are compatible with the corresponding standardized model classes and postprocessing rules.

By registering prediction pipelines, MultiMolecule turns standardized checkpoints into deployment-facing tools.
A user can apply a pretrained checkpoint directly, export a Runner-fine-tuned checkpoint into a compatible model directory and then use the same task interface for inference.
The resulting predictions are traceable to the model implementation, standardized checkpoint, tokenizer, postprocessing rule and associated validation evidence.
This pipeline layer extends MultiMolecule from a training and evaluation ecosystem to a practical biological prediction Resource.
The next section summarizes how Extended Data tables and validation outputs tie these implementations, checkpoints, datasets, workflows and pipelines to public validation evidence.

\subsection{Validation checks and Extended Data summaries connect Resource claims to public evidence}\label{subsec:validation-evidence}

Validation evidence is reported at the same levels at which users encounter the Resource.
MultiMolecule therefore reports checks for model-family implementations, standardized model checkpoints, dataset resources, Runner workflows and prediction pipelines.
For each public asset, the manuscript links the fields relevant to that asset class, including source provenance, public identifier, conversion or preparation procedure, tested quantities, validation criterion, software environment and documentation route.
These checks and summaries make the Resource inspectable as a scientific release, with each claim traceable to an Extended Data table, source-tree script, model or dataset card, example workflow or manuscript validation output.
The corresponding model-family, checkpoint, dataset-resource and prediction-pipeline summaries are provided in Extended Data Tables~\ref{tab:ed-completeness}--\ref{tab:ed-pipelines}.
Runner-workflow evidence is reported through the representative workflow artifacts, example outputs and validation checks described above rather than as a separate Extended Data table.

The validation evidence is organized around the reusable layers of the Resource.
Implementation summaries document model-facing completeness.
Checkpoint-conversion summaries document source-versus-MultiMolecule agreement during standardization, covering 112 standardized checkpoints and 177 output-level checks in the manuscript-reported inventory.
Tokenizer and sequence-I/O tests document vocabulary, special-token and biological-token alignment.
Dataset summaries document provenance, schemas and split policies.
Runner examples document executable workflows.
Pipeline tests document accepted inputs, compatible checkpoints, returned fields and deterministic postprocessing.
Together, these checks and summaries define what has been checked for the manuscript-reported inventory and where each Resource claim is anchored.

This validation approach defines how the Resource can grow without changing its evidentiary standard.
A model family, standardized checkpoint, dataset resource, Runner workflow or prediction pipeline enters the Resource only when it can be linked to its implementation, conversion or preparation procedure, execution or loading interface, validation evidence and public documentation.
New components therefore enter through the same completeness, conversion, workflow, deployment and documentation requirements.
This reporting structure supports MultiMolecule as a maintainable community Resource.
The manuscript counts refer to the manuscript-reported Resource state summarized in the Results and Extended Data, while the public repository, package distribution, Hub assets and documentation remain maintained as continuing access points.
This keeps future additions reviewable against the same public evidence standard without treating the ecosystem as a static catalogue.

\section{Discussion}\label{sec:discussion}

MultiMolecule shifts the reusable unit of biomolecular sequence modeling from a checkpoint entry or repository wrapper to a source-checked model-family implementation linked to standardized checkpoints, executable workflows, prediction interfaces and validation evidence.
This shift addresses a practical bottleneck in a field where useful models are abundant, but reusable model releases remain difficult to preserve, evaluate, adapt and deploy across laboratories.
A model family is counted only when its model-facing interface, task heads, checkpoint conversion, workflow compatibility and documentation enter the same Resource conventions.
The resulting ecosystem is curated for executable completeness: each counted implementation is working, source-checked and documented.

This organization makes model reuse a scientific workflow rather than a reconstruction exercise.
In the absence of standardized implementations, users must often recover tokenizer conventions, vocabulary order, checkpoint names, tensor layouts, output objects, task heads and preprocessing assumptions from source repositories.
These details can affect model behavior and downstream comparisons, yet they are rarely visible in high-level performance summaries.
By absorbing such conventions into conversion scripts, model cards, validation checks and shared interfaces, MultiMolecule makes the assumptions surrounding a model inspectable.
The conversion checks and their Extended Data summaries are therefore part of the Resource infrastructure: they document what was standardized, what was compared and what behavior is expected to be preserved before a standardized checkpoint is used for controlled evaluation, fine-tuning or deployment.

MultiMolecule complements, rather than replaces, existing infrastructure for computational biology.
Tensor libraries and model hubs provide essential primitives for computation and checkpoint distribution \cite{paszke2019pytorch,wolf2020transformers}.
Workflow systems improve provenance and portability for multi-step analyses \cite{wratten2021reproducible}.
Scalable biomolecular AI frameworks emphasize distributed training and high-throughput model development \cite{stjohn2024bionemo}.
Faithful single-architecture reimplementations can deeply reproduce one influential model family \cite{ahdritz2024openfold}.
Earlier genomics model zoos established the value of shared access to trained predictors \cite{avsec2019kipoi}.
MultiMolecule builds on these ideas by standardizing the domain-specific layer that connects biological sequence I/O, molecular tokenization, complete model-family implementations, standardized model checkpoints, reusable modules, Runner workflows, prediction pipelines and validation evidence across heterogeneous biomolecular model families.
In this sense, MultiMolecule follows the model-hub and model-zoo tradition, but shifts the reusable unit from a downloadable checkpoint to a source-behavior-preserving implementation and workflow layer.
Earlier genomics model zoos often preserved model-specific execution environments to make trained predictors accessible.
MultiMolecule takes a complementary approach by reimplementing model-facing behavior natively against a shared environment and interface, allowing model families to interoperate within one installation when their redistribution terms permit standardization.
It distributes the resulting standardized checkpoints and processed datasets through the Hugging Face Hub rather than a bespoke catalogue, adding the shared implementation, workflow and validation layer over the checkpoint-distribution primitives that hubs already provide.

The breadth of the Resource reflects a deliberately neutral view of biomolecular modeling.
Task-specific predictors, conventional deep-learning architectures and pretrained sequence backbones remain useful for different biological questions, and no single modeling paradigm should determine how sequence models are reused.
MultiMolecule therefore treats these model types as first-class families under the same implementation and validation requirements.
This choice is important for RNA, transcript, genomic, splicing, regulatory and protein workflows, where older task-specific architectures and newer pretrained models often coexist.
The Resource helps users select model assets by biological task, interface compatibility and available validation evidence.

Several limitations define the Resource state reported here.
Conversion checks compare standardized checkpoints with deterministic source-reference outputs and report residual numerical differences---from hardware, precision settings, dependency versions or undocumented source behavior---as tolerances and validation criteria.
Coverage is broad but bounded: a family, checkpoint, dataset or pipeline enters the Resource once it can be linked to implementation or preparation scripts, validation evidence and documentation, so some scientifically important assets remain outside it when their model weights, conventions, source code or redistribution terms do not support public standardization.
Dataset-to-workflow wiring relies on declared or inferable schemas, and ambiguous labels require user declaration.
Curated dataset resources are documented processed datasets; a leakage-free comparison holds only where a split policy provides one.
Backend-performance characterization across architecture classes is outside the central Resource claims of this manuscript.
MultiMolecule is distributed under AGPL-3.0-or-later as a reciprocal open-resource release; license scope and redistribution policy are specified in Methods and Code availability.

Resource sustainability depends on preserving the same evidentiary standard as the Resource grows.
Public model and dataset cards, conversion scripts, Extended Data summaries, Hugging Face assets and Zenodo metadata keep the Resource inspectable as it evolves.
Future additions can therefore be reviewed against public documentation rather than informal maintainer knowledge: new model families should enter as complete implementations, new standardized checkpoints should enter with conversion summaries, new datasets should enter with schema and provenance documentation, new workflows should enter with executable artifacts and new pipelines should enter with registered task interfaces and postprocessing documentation.
This structure allows the Resource to expand without changing its conceptual boundary.
By organizing biomolecular sequence models as complete, source-checked and executable implementations linked to standardized checkpoints, MultiMolecule lets biomolecular models be preserved, inspected, adapted and deployed as reusable scientific infrastructure.

\section{Methods}\label{sec:methods}

\subsection{Software architecture and package organization}\label{subsec:methods-architecture}

MultiMolecule is implemented as a Python ecosystem whose public interfaces are organized around biological sequence I/O, molecular tokenization, model-family implementations, reusable modules, Runner execution and prediction pipelines.
These interfaces correspond to the assets counted in the manuscript-reported inventory: complete model-family implementations, standardized model checkpoints, curated dataset resources, Runner workflows and prediction-pipeline registry entries.
The main user-facing entry points are standardized model loading through \texttt{AutoModel} interfaces, configuration-driven execution through the \texttt{mmtrain}, \texttt{mmevaluate} and \texttt{mminfer} commands, checkpoint conversion through per-family \texttt{convert\_checkpoint.py} scripts and deployment through registered prediction-pipeline task names.
Supporting utilities are used by these interfaces but are not treated as separate Resource claims in this manuscript.

\subsection{Resource scope and inclusion policy}\label{subsec:methods-release-scope}

MultiMolecule includes model families, checkpoint variants, dataset resources and pipelines according to a Resource-level inclusion policy.
MultiMolecule distributes framework code, conversion scripts, standardized model checkpoints and processed datasets under AGPL-3.0-or-later.
A model family, checkpoint variant or dataset resource is included only when its public implementation details, tokenizer or preprocessing conventions, redistributable checkpoints or processed datasets and provenance information allow MultiMolecule to redistribute the standardized implementation, checkpoint or processed dataset, together with conversion or preparation scripts and documentation, under AGPL-3.0-or-later.
Included assets are accompanied by validation evidence for the standardized asset; where upstream redistribution terms are incompatible with AGPL-3.0-or-later redistribution by MultiMolecule, the asset is excluded rather than shipped as an unverifiable wrapper.
This describes the license of the MultiMolecule distribution and does not assert that upstream sources adopt the same license.
The policy narrows the catalogue but keeps the Resource inspectable and complete: users receive not only framework code but the associated checkpoints, processed datasets, conversion scripts and documented validation evidence.

The reported Resource inventory describes the manuscript-reported Resource state summarized in the Results and Extended Data.
MultiMolecule is maintained as a live Resource: public code, package distribution, standardized checkpoints, dataset assets, documentation and citable release metadata are served through coordinated access points that may include subsequent additions, corrections or deprecations.
The Extended Data tables define the inventory underlying the manuscript counts, whereas the public access points provide maintained routes to the live Resource.
Dataset resources are included when a conversion definition, source provenance, processed fields and dataset-card metadata are available and when the resulting datasets can be publicly redistributed by MultiMolecule under AGPL-3.0-or-later.
Public model and dataset repository counts may exceed source-tree family or resource counts because a single implementation or conversion definition can produce multiple released checkpoints, variants or processed datasets.

\subsection{Completeness criteria for model-family implementations}\label{subsec:methods-completeness}

A model-family implementation is considered complete in the manuscript-reported inventory when it satisfies five criteria.
First, the in-tree implementation reproduces the upstream model-facing public classes and task heads exposed by the source release, excluding repository-specific training harnesses, dataset loaders, visualization utilities or analysis scripts that are not part of the model API.
Second, the implementation registers standardized configuration, tokenizer, model-output and AutoModel interfaces under MultiMolecule conventions.
Third, the implementation provides checkpoint-conversion code for every redistributable checkpoint variant included in the manuscript-reported inventory.
Fourth, converted checkpoints are evaluated against deterministic source-reference outputs over the source-exposed quantities selected for that family, including hidden states, logits, contact maps, pooler outputs or task-head outputs when available.
Fifth, the implementation exposes the standardized model interface required for Runner workflows, allowing it to participate in configured sequence-level, token-level or contact-level tasks, and is validated for user-facing pipeline compatibility when the family supports a registered prediction pipeline.
Extended Data summaries document the criteria satisfied by each family, the corresponding conversion and conversion-validation scripts, tested outputs, numerical tolerances and validation status.

\subsection{Model standardization and checkpoint conversion}\label{subsec:methods-conversion}

Each supported model family is organized as an implementation containing configuration classes, model classes, task-specific public heads, checkpoint-conversion code, tokenizer registration and model-card documentation.
Model implementations are independently reimplemented under shared interfaces rather than copied from upstream repositories.
MultiMolecule uses checkpoint-conversion scripts to produce standardized model checkpoints from redistributable source checkpoints.
Checkpoint-conversion scripts translate upstream state-dict names, tensor layouts, vocabulary order, special-token conventions and architecture-specific parameters into the MultiMolecule representation.
Converted checkpoints are saved with the corresponding configuration, tokenizer files, model weights and model-card metadata.
Tokenizer remapping is handled during checkpoint conversion: vocabulary order, special-token conventions and upstream-to-MultiMolecule token mappings are encoded in the converted checkpoint and tokenizer files so that runtime inference uses the standardized MultiMolecule representation.
Token-ID equality with upstream tokenizers is therefore not used as a general validation criterion.

Checkpoint conversion is technically nontrivial because upstream model releases differ along several independent axes that are not visible at the level of weight tensors alone.
Three representative cases illustrate why a uniform conversion protocol, rather than a checkpoint-format adapter, is required.

\emph{Case 1: vocabulary reordering and codon variants in RNA-FM.}
The published RNA-FM checkpoint uses a 25-entry single-letter RNA vocabulary, and the companion mRNA-FM checkpoint uses a 73-entry 3-mer codon vocabulary.
Both checkpoints place special tokens at upstream-specific positions and pad the remaining slots.
The MultiMolecule converter resolves both variants in a single \texttt{convert\_checkpoint.py} (\texttt{multimolecule/models/rnafm/}) by selecting the alphabet from the checkpoint filename, reordering the rows of \texttt{word\_embeddings} and \texttt{lm\_head.decoder} to MultiMolecule order via \texttt{convert\_word\_embeddings()}, and writing a tokenizer configuration that stores the codon flag where applicable.
Without this remapping, callers would have to remap token IDs at every forward pass.

\emph{Case 2: state-dict path renames in models reused across architectures.}
RNA-FM, RiNALMo and several RNA-BERT variants share a transformer encoder layout but ship state-dict keys under incompatible upstream prefixes: RNA-FM uses ESM-style names such as \texttt{self\_attn} and \texttt{fc1}/\texttt{fc2}, RiNALMo names the same submodules \texttt{mh\_attn} and \texttt{transition}, and RNA-BERT nests attention under \texttt{selfattn}.
The MultiMolecule converters apply a sequence of regular-expression rewrites that translate these prefixes into the standardized \texttt{model.embeddings.*} and \texttt{model.encoder.layer.*.attention.*} naming used across MultiMolecule implementations, and additionally translate \texttt{LayerNorm} / \texttt{gamma} / \texttt{beta} conventions into \texttt{layer\_norm} / \texttt{weight} / \texttt{bias}.
This makes module-level APIs consistent across families that share a transformer encoder backbone, even when their upstream releases name the same parameters differently.

\emph{Case 3: tied weights and head-specific layouts in RNA-FM.}
When \texttt{tie\_word\_embeddings} is set, the language-modeling decoder reuses \texttt{word\_embeddings} weights; the converter must rewrite \texttt{lm\_head.decoder.weight} to the reordered word embeddings rather than the upstream decoder tensor.
RNA-FM's pairwise contact head additionally ships convolutional kernels under upstream-specific names (\texttt{proj.first.0}, \texttt{proj.resnet}, \texttt{proj.final}); the MultiMolecule converter remaps these into the standardized \texttt{ss\_head.projection}, \texttt{ss\_head.convnet.layers.<idx>} and \texttt{ss\_head.prediction} paths, drops the superseded upstream decoder tensors once the standardized prediction head is present, and validates the resulting weights against deterministic source-reference outputs.
These cases are illustrative rather than exhaustive; analogous per-family details are encoded in each model's \texttt{convert\_checkpoint.py} and checked by the conversion-validation workflow described in the next section.

\subsection{Source-reference generation and conversion checking}\label{subsec:methods-golden}

Source-model reference outputs are generated separately from the MultiMolecule conversion code.
We refer to these source-reference outputs as \emph{golden outputs} in the committed fixtures to match standard software-testing terminology; they are reference-implementation outputs, not biological ground truth.
The \texttt{multimolecule/upstream} repository contains source-adapted execution scripts that run original model implementations on deterministic reference inputs.
These scripts generate source-reference outputs without importing the standardized MultiMolecule model implementations.
Committed fixture cases are stored in the \texttt{multimolecule/golden} repository as plain directories containing \texttt{inputs.safetensors}, \texttt{expected.safetensors} and \texttt{meta.json}.
The fixture metadata store source identifiers, input metadata, output-channel metadata, tolerances and provenance fields.
The main MultiMolecule repository consumes these fixtures through the stable layout \texttt{models/<model>/<checkpoint-id>/}.

During checkpoint conversion, each \texttt{convert\_checkpoint.py} script maps source checkpoint parameters, tokenizer files and vocabulary conventions into the standardized MultiMolecule representation.
The corresponding conversion-validation workflow evaluates the standardized implementation on the same deterministic inputs and compares source-exposed quantities with the matching source-reference outputs.
Each check uses deterministic reference inputs chosen to exercise the model vocabulary and the source-exposed output paths selected for that family.
Because conversion is a structural remapping---vocabulary reordering, parameter renaming and tied-weight resolution---these inputs test whether the converted weights, tokenizer mapping and standardized implementation reproduce selected source-reference outputs through the forward computation.
They do not characterize all length-dependent, batch-dependent or backend-dependent behaviors.
The retained manuscript validation matrix therefore reports conversion agreement on fixed deterministic reference inputs rather than serving as a general numerical-behavior benchmark.
Depending on the source family, tested quantities include hidden states, logits, contact maps, pooler outputs, vocabulary embeddings or task-head predictions.
When exact equality is not expected because of floating-point precision, hardware differences or dependency-version effects, the conversion summary reports the comparison metric and tolerance rather than treating them as implicit implementation details.
Because these numerical differences accumulate with model depth and width, the tolerance is reported per checkpoint rather than as a single global threshold: many small or shallow conversions reproduce their source references exactly, whereas the largest checkpoints, such as the biggest protein language models, require correspondingly wider bounds.

For each tested output of each standardized checkpoint in the manuscript comparison matrix, the conversion summary reports the standardized checkpoint identifier and absolute-difference statistics for that output.
The manuscript comparison outputs are generated from the checkpoint-level and output-level comparison summaries and are reported in Extended Data Table~\ref{tab:ed-checkpoints}.
Checkpoint-level status is therefore traceable to output-wise comparisons rather than reported only as a binary result.
This separation makes the reference-generation scripts, committed source-reference fixtures, conversion scripts and Extended Data summaries inspectable without conflating fixture generation with checkpoint conversion.

\subsection{Dataset parsing and task inference}\label{subsec:methods-task-inference}

Datasets are loaded from local files, tabular objects or public dataset repositories and converted to an internal table representation.
The dataset layer identifies identifier columns, biological sequence columns, secondary-structure columns, feature columns and label columns using column names, string values and molecular alphabet recognition.
For label columns, task level is inferred from label geometry relative to the biological sequence length.
Whole-sequence labels are assigned to sequence-level prediction, per-position labels to token-level prediction and pairwise matrices to contact-level prediction.
Task type is inferred from label data type, label dimensionality and class cardinality, distinguishing binary classification, multiclass classification, multilabel classification and regression.
When tokenizers introduce special tokens or when sequence truncation is applied, label alignment is computed against the biological sequence length so that token-level and contact-level labels remain biologically meaningful.
Users can override inferred feature columns, label columns, sequence columns, ignored columns, discrete label maps and task definitions when dataset-specific conventions are ambiguous.
The task-inference decision process first identifies candidate sequence and identifier columns, then assigns label columns to sequence-level, token-level or contact-level tasks according to label geometry relative to biological sequence length.
Task type is then assigned from data type, dimensionality and class cardinality.
Validation examples include curated cases with known task definitions, synthetic edge cases for missing values and ambiguous columns, and examples requiring explicit user override.

For each curated dataset resource, the dataset summary stores the upstream source, preprocessing script, processed fields, sequence column, label schema and redistribution status.
Dataset resources are treated as documented processed datasets.
They are not treated as automatic leakage-free benchmarks for arbitrary cross-dataset comparisons unless a split policy explicitly provides that guarantee.
Dataset cards and preprocessing documentation therefore describe how the processed dataset was constructed and how its fields should be interpreted, rather than implying that every downstream comparison using it is a validated benchmark.

The manuscript-reported Resource inventory includes 16 curated dataset resources.
Each is converted from one or more upstream sources, exposes a documented label schema and is distributed through the MultiMolecule Hugging Face Hub organization.
Full dataset-resource summaries are provided in Extended Data Table~\ref{tab:ed-datasets}.

\subsection{Runner configuration and execution}\label{subsec:methods-runner}

Runner workflows are specified by configuration files that define the data source, split structure, pretrained checkpoint or newly initialized backbone, optional necks and heads, data-loader settings, optimizer, scheduler, checkpoint restoration, random seed and execution mode.
At initialization, the Runner resolves the tokenizer, builds datasets, constructs task descriptors and instantiates the standardized model family through the module registry.
The resulting workflow is assembled from a standardized model family and task-defined execution components rather than from a source-repository training script.
Task descriptors inferred from, or declared for, the dataset specify the prediction level, target dimensionality, objective, metrics and output schema.
This design makes model-task pairing a configuration-level operation: the model family supplies the standardized forward interface, whereas the task descriptor defines how dataset labels are interpreted and evaluated.
During training, the Runner moves batches to the selected device, executes the model, aggregates task-specific losses, updates metrics and applies optimization steps.
During evaluation, it restores the requested checkpoint when needed and applies the same task-aware model without parameter updates.
During inference, it collects model predictions and matched labels when labels are present.
For pretrained backbones, parameter groups can assign different learning-rate and weight-decay settings to pretrained and newly initialized components.
For each reported Runner workflow, the validation evidence includes the resolved configuration file, command-line entry point, tokenizer, standardized checkpoint, expected task descriptor, instantiated task-specific components, metrics and output files.
The Runner supports sequence-level, token-level, contact-level and inference-only execution paths, selected through configuration.
Downstream studies that use MultiMolecule implementations under independent (non-Runner) training stacks are treated as model-layer reuse evidence, not as Runner-validation evidence.
Runner-validation evidence is reserved for workflows executed through the MultiMolecule Runner.
This distinction prevents model-implementation reuse from being conflated with validation of the Runner execution layer.

\subsection{Prediction-pipeline registry and tests}\label{subsec:methods-pipelines}

Each registered prediction pipeline specifies accepted biological inputs, compatible standardized checkpoints, tokenizer requirements, batching behavior, postprocessing rules and returned output fields.
Pipeline registry entries distinguish user-facing prediction tasks from the internal task descriptors used by the Runner.
The user-facing task name defines the registered task interface, whereas internal task descriptors define how dataset labels are wired to heads, criteria, metrics and output schemas during training or evaluation.
Pipeline validation checks model loading, input validation, output-field structure, deterministic postprocessing and error behavior for incompatible inputs or checkpoints.
Hosted default-checkpoint smoke tests are used for deployment monitoring, but all-pass smoke status is not treated as a display-level result.

\subsection{Validation checks and reporting}\label{subsec:methods-validation}

Quality-control analyses are organized around the claims made by the Resource; they validate Resource functionality and do not attempt to rank biological performance across models.
Tokenizer tests evaluate standardized vocabulary definitions, special-token behavior, encoding and decoding invariants, and documented upstream-to-MultiMolecule vocabulary mappings when conversion requires remapping.
Token-ID equality with upstream tokenizers is not expected for many language-model checkpoints because MultiMolecule may remap or standardize tokenizer conventions during checkpoint conversion.
Checkpoint-conversion checks evaluate numerical agreement between converted checkpoints and source-reference fixtures generated from source implementations on deterministic reference inputs, reporting element-wise absolute-difference summaries, including minimum, 25th percentile, median, 75th percentile, maximum and mean differences, against combined absolute and relative tolerances, with exact-match status where applicable, according to the per-block or task-specific quantity being tested.
Task inference is validated against the declared task definitions of the curated dataset resources and on synthetic edge cases spanning sequence-level, token-level and contact-level labels; datasets without a declared task receive the inferred assignment, and cases where the schema is ambiguous require user override and are documented separately.
Runner quality-control checks test the task-defined workflow path rather than only individual example scripts.
These checks cover configuration parsing, tokenizer resolution, dataset construction, task-descriptor construction, model instantiation, task-specific execution, metric computation, forward execution and output writing across sequence-level, token-level, contact-level and inference-only execution modes.
Full training, evaluation and inference are further exercised by full-stack workflows such as CHANRG and Uni-RNA secondary-structure training, which use the MultiMolecule Runner under structured-prediction settings.
This reporting distinguishes execution-level model-task composition from downstream biological performance benchmarking.
Pipeline tests evaluate input handling, compatible-checkpoint loading, output formatting and biological postprocessing for deployment-facing workflows; pipeline reporting includes accepted sequence inputs, compatible checkpoint requirements, returned output fields, deterministic postprocessing choices and edge cases that require user interpretation.
For the RNA secondary-structure pipeline specifically, these tests include contact-probability or contact-map output fields and dot-bracket output behavior when requested.
Extended Data summaries and manuscript validation files report dataset and model panels, tested quantities, tolerances and pass/fail criteria.
Environment details are reported only for analyses for which they are needed and were collected.

\subsection{Hardware and software environment}\label{subsec:methods-environment}

Reproducibility metadata are reported where they were collected and are needed to interpret a specific artifact.
Upstream fixture generation is documented in the \texttt{multimolecule/upstream} repository through generator scripts, source metadata and Docker or dependency specifications where they are needed.
The main package environment is described by the repository and package metadata.
Where numerical variation is possible, conversion summaries report explicit tolerances and comparison criteria.

\subsection{Availability, release metadata and reproducibility}\label{subsec:methods-availability}

The source code, documentation, package releases, model checkpoints and dataset assets are maintained through coordinated public access points.
The repository and package metadata provide the documentation site, package installation route, repository URL, license, supported Python versions and core dependency requirements.
Model checkpoints and dataset assets are distributed through the MultiMolecule Hugging Face Hub organization, and citable release metadata are provided through Zenodo.
Model cards and dataset cards document source provenance, intended use, task or usage notes and citation information.
The manuscript counts describe the manuscript-reported Resource state, while these access points provide the maintained routes for code, checkpoints, dataset assets, documentation and citation metadata.

\subsection{Maintenance and contribution policy}\label{subsec:methods-maintenance}

Resource maintenance and community contributions follow documented requirements designed to keep Resource assets, documentation and validation expectations consistent across versions.
New model families are considered for inclusion when public implementation details, tokenizer definitions, checkpoints and redistribution terms allow behavior-preserving reimplementation, checkpoint conversion, conversion-validation checking and release by MultiMolecule under AGPL-3.0-or-later; an in-tree \texttt{convert\_checkpoint.py}, a conversion-validation script and a model card are required for each accepted family.
New dataset resources are considered when source provenance, preprocessing rules, label schema and redistribution terms can be documented in a dataset card and the resulting processed datasets can be redistributed by MultiMolecule under AGPL-3.0-or-later.
Behavior-preserving changes within MultiMolecule's reimplemented model files include reconciliations of implementation-specific behaviors, as well as conversion adaptations required by the standardized model format.
Behavior-affecting corrections that can be applied independently of MultiMolecule's reimplementation format are communicated to original maintainers as upstream GitHub issues or pull requests; format-tied conversion adaptations are kept within the Resource and documented in the relevant model card or conversion code.
Issues, conversion-validation failures or upstream license changes that make a Resource asset incompatible with distribution by MultiMolecule under AGPL-3.0-or-later may lead to deprecation in a future Resource update.
Issue and pull-request pathways, model-card and dataset-card requirements, and validation expectations are documented alongside the codebase so that contributions and downstream forks can be reviewed against the same criteria.

\pagebreak

\backmatter

\bmhead{Supplementary information}

No separate Supplementary Information file accompanies this manuscript.
No separate Supplementary Information tables are provided beyond the Extended Data tables included with this manuscript.

\bmhead{Acknowledgements}

The author would like to acknowledge Chang Liu.
This work was made over years in which much remained unfinished: the code, the paper and the life around them.
Through those years, she gave me what no public artifact can preserve: patience, constancy and support.
We met as students, and walk on together still.
Everything public in this paper rests on something private and immeasurable: the love by which I learned to continue.
Thank you, Chang, for every year, and for every beginning.



\section*{Declarations}

\bmhead{Funding}

The author received no external funding for this work.

\bmhead{Competing interests}

The author declares no competing interests.

\bmhead{Ethics approval and consent to participate}

Not applicable.

\bmhead{Consent for publication}

Not applicable.

\bmhead{Data availability}

Dataset resources used by MultiMolecule are available through the MultiMolecule Hugging Face Hub organization (\url{https://huggingface.co/multimolecule}) and are documented in the corresponding dataset cards.
Standardized model checkpoints are available through the MultiMolecule Hugging Face Hub organization and are documented in the corresponding model cards.

\bmhead{Materials availability}

Not applicable.

\bmhead{Code availability}

The MultiMolecule source code is available at \url{https://github.com/MultiMolecule/multimolecule} under AGPL-3.0-or-later.
Source-reference generation scripts and committed source-reference fixtures are available at \url{https://github.com/MultiMolecule/upstream} and \url{https://github.com/MultiMolecule/golden}, respectively.
Framework code, conversion scripts, standardized model checkpoints and processed datasets distributed by MultiMolecule are released under AGPL-3.0-or-later.
Documentation is available at \url{https://multimolecule.danling.org}.
The software is distributed as an installable Python package at \url{https://pypi.org/project/multimolecule}, with standardized model checkpoints and dataset assets available at \url{https://huggingface.co/multimolecule}.
Citable release metadata are deposited on Zenodo (DOI \url{https://doi.org/10.5281/zenodo.15119050}).

\bmhead{Author contributions}

Z.C.\ is the sole author and is responsible for the conception, implementation, validation, drafting and revision of this work.

\begin{appendices}
  \setcounter{table}{0}
  \renewcommand{\thetable}{\arabic{table}}
  \renewcommand{\tablename}{Extended Data Table}

  \section{Extended Data 1: Model-family implementations}\label{secA1}

  This table provides a per-family overview of the complete model-family implementations counted in the manuscript-reported inventory, listing each family's molecule type, architecture, parameter size, input tokenization, maximum context length and standardized-checkpoint count.
  It is generated from the manuscript result files and source-tree metadata rather than hand-maintained prose.

  \begin{landscape}
    
\begingroup
\tiny
\setlength{\tabcolsep}{1.5pt}
\emergencystretch=2em
\begin{longtable}{@{}p{0.13\linewidth}p{0.05\linewidth}p{0.045\linewidth}p{0.25\linewidth}p{0.10\linewidth}p{0.105\linewidth}p{0.075\linewidth}p{0.03\linewidth}@{}}
\caption{\textbf{Model-family implementation matrix.} Each row is a model-family implementation counted in the manuscript-reported inventory, summarizing its molecule type, architecture class, parameter size (a range when the family's standardized checkpoints differ in size), input tokenization, maximum context length (\texttt{var.} where the architecture imposes no fixed limit) and number of standardized checkpoints (N). Every counted family additionally ships a model card, checkpoint converter, configuration and modeling files, automatic-loading registration and a task-specific public class; completeness is evaluated at the model-facing interface level rather than at the full source-repository level. The Ref column cites the family's original publication.}\label{tab:ed-completeness}\\
\toprule
Family & Ref & Mol. & Architecture & Size & Tokenization & Max len & N \\
\midrule
\endfirsthead
\toprule
Family & Ref & Mol. & Architecture & Size & Tokenization & Max len & N \\
\midrule
\endhead
a2z-chromatin & \cite{wrightsman2022a2zchromatin} & dna & CNN + BiLSTM (DanQ) & 1.23M & nucleotide & 600 & 2 \\
AbLang & \cite{olsen2022ablang} & protein & Transformer encoder (BERT-style) & 85.8M & amino-acid & 160 & 2 \\
AbLang2 & \cite{olsen2024ablang2} & protein & Transformer encoder (RoPE + SwiGLU) & 44.8M & amino-acid & 256 & 1 \\
AMPLIFY & \cite{fournier2024amplify} & protein & Transformer encoder (RMSNorm, SwiGLU and RoPE) & 120M-350M & amino-acid & 2,048 & 2 \\
APARENT & \cite{bogard2019aparent} & rna & 1D CNN (conv + dense) & 6.43M & nucleotide & 205 & 1 \\
APARENT2 & \cite{linder2022aparent2} & rna & Residual CNN (ResNet) & 0.19M & nucleotide & 205 & 1 \\
Basenji & \cite{kelley2018basenji} & dna & 1D dilated residual CNN & 30M & nucleotide & 131,072 & 1 \\
Basset & \cite{kelley2016basset} & dna & Convolutional neural network (1D CNN) & 4.14M & nucleotide & 600 & 1 \\
Borzoi & \cite{linder2025borzoi} & dna & Conv stem + Transformer + U-Net (Enformer-style) & \textasciitilde{}185M & nucleotide & 524,288 & 2 \\
BPfold & \cite{zhu2025bpfold} & rna & Transformer encoder + pairwise CNN & 47.8M & nucleotide & 600 & 1 \\
BPNet & \cite{avsec2021bpnet} & dna & Dilated CNN (residual, profile+count heads) & 0.13M & nucleotide & 1,000 & 1 \\
CaLM & \cite{outeiral2022calm} & dna & Transformer encoder (ESM- or BERT-style) & 85.8M & codon & 1,024 & 1 \\
CARP & \cite{yang2024carp} & protein & Dilated CNN (ByteNet) & 0.61M-643M & amino-acid & 1,024 & 4 \\
ChromBPNet & \cite{pampari2024chrombpnet} & dna & Dilated CNN (BPNet-style) & 6.61M & nucleotide & 2,114 & 1 \\
DeepCpG-DNA & \cite{angermueller2017deepcpgdna} & dna & 1D CNN (DeepCpG DNA submodule) & 4.11M-4.43M & nucleotide & 1,001 & 5 \\
DeepMEL & \cite{minnoye2020deepmel} & dna & CNN + BiLSTM & 3.44M & nucleotide & 500 & 1 \\
DeepSEA & \cite{zhou2015deepsea} & dna & 1D CNN (DeepSEA) & 52.8M & nucleotide & 1,000 & 1 \\
DeepSTARR & \cite{dealmeida2022deepstarr} & dna & 1D CNN (DeepSTARR) & 0.62M & nucleotide & 249 & 1 \\
DeltaSplice & \cite{xu2024deltasplice} & rna & Dilated residual 1D CNN (ensemble) & 40.4M & nucleotide & var. & 2 \\
DNABERT & \cite{ji2021dnabert} & dna & Transformer encoder (BERT-style) & 86M-89M & 3-6-mer & 512 & 4 \\
DNABERT-2 & \cite{zhou2024dnabert2} & dna & Transformer encoder (MosaicBERT, ALiBi) & 117M & BPE & 512 & 1 \\
DNABERT-S & \cite{zhou2025dnaberts} & dna & Transformer encoder (MosaicBERT with ALiBi) & 117M & BPE & 512 & 1 \\
Enformer & \cite{avsec2021enformer} & dna & Conv stem + Transformer trunk & 246M & nucleotide & 196,608 & 1 \\
ERNIE-RNA & \cite{yin2024ernierna} & rna & Transformer encoder (BERT-style) & 85.7M & nucleotide & 1,024 & 2 \\
Framepool & \cite{karollus2021framepool} & rna & 1D residual CNN (frame-aware pooling) & 0.28M & nucleotide & var. & 1 \\
HAL & \cite{rosenberg2015hal} & rna & Linear model (hexamer k-mer regression) & \textasciitilde{}4K & nucleotide & 160 & 1 \\
HyenaDNA & \cite{nguyen2023hyenadna} & dna & Hyena operator (implicit conv, decoder-only) & 0.45M-6.62M & nucleotide & 1,000,002 & 4 \\
Malinois & \cite{gosai2024malinois} & dna & 1D CNN (Basset-style branched) & 4.11M & nucleotide & 600 & 1 \\
MaxEntScan & \cite{yeo2004maxentscan} & rna & Maximum-entropy scorer (table lookup) & n/a (table) & nucleotide & 9 & 2 \\
MMSplice & \cite{cheng2019mmsplice} & rna & Modular 1D CNN (5 sub-networks) & 57K & nucleotide & var. & 1 \\
MPRA-DragoNN & \cite{movva2019mpradragonn} & dna & 1D CNN (DragoNN ConvModel) & 0.34M & nucleotide & 145 & 1 \\
MTSplice & \cite{cheng2021mtsplice} & rna & Dilated CNN (dual-tower) & 0.21M & nucleotide & 800 & 1 \\
OpenSpliceAI & \cite{chao2025openspliceai} & rna & Dilated residual CNN (SpliceAI-style) & 0.09M-0.7M & nucleotide & var. & 20 \\
Optimus 5-Prime & \cite{sample2019optimus5prime} & rna & 1D CNN (regression) & 0.48M & nucleotide & 50 & 1 \\
OptMRL & \cite{korbel2024optmrl} & rna & 1D CNN (3 conv + dense) & 0.48M & nucleotide & 50 & 1 \\
Pangolin & \cite{zeng2022pangolin} & rna & Dilated residual 1D CNN (SpliceAI-style) & 8.36M & nucleotide & var. & 1 \\
ProCapNet & \cite{cochran2024procapnet} & dna & Dilated CNN (BPNet-style) & 6.43M & nucleotide & 2,114 & 1 \\
ProGen2 & \cite{nijkamp2023progen2} & protein & GPT-J-style autoregressive Transformer & 151M-6.4B & amino-acid & 2,048 & 7 \\
ProteinBERT & \cite{brandes2022proteinbert} & protein & Conv + global-attention (ProteinBERT) & 16M & amino-acid & 1,024 & 1 \\
RibonanzaNet & \cite{he2024ribonanzanet} & rna & Transformer encoder + 1D conv (BERT-style) & 11.4M & nucleotide & var. & 4 \\
RiNALMo & \cite{penic2025rinalmo} & rna & Transformer encoder (BERT-style, RoPE) & 30M-650M & nucleotide & 1,024 & 4 \\
RNABERT & \cite{akiyama2022rnabert} & rna & Transformer encoder (BERT-style) & 0.48M & nucleotide & 440 & 1 \\
RNAErnie & \cite{wang2024rnaernie} & rna & Transformer encoder (BERT- or ERNIE-style) & 86.1M & nucleotide & 512 & 1 \\
RNA-FM & \cite{chen2022rnafm} & rna & Transformer encoder (BERT- or ESM-style) & 99.5M-239M & nucleotide or codon & 1,024 & 3 \\
RNA-MSM & \cite{zhang2023rnamsm} & rna & MSA Transformer (axial attention) & 95.9M & nucleotide & 1,024 & 1 \\
scBasset & \cite{yuan2022scbasset} & dna & 1D CNN (Basset-style) & 4.59M & nucleotide & 1,344 & 1 \\
SpliceBERT & \cite{chen2023splicebert} & rna & Transformer encoder (BERT-style) & 19.4M-19.7M & nucleotide & 1,024 & 3 \\
SPOT-RNA & \cite{singh2019spotrna} & rna & Ensemble 2D CNN (+BiLSTM/dilated) & 17.5M & nucleotide & var. & 1 \\
SpTransformer & \cite{you2024sptransformer} & rna & CNN + Sinkhorn Transformer & 17.1M & nucleotide & 8,192 & 1 \\
UFold & \cite{fu2022ufold} & rna & U-Net (encoder-decoder CNN) & 8.64M & nucleotide & var. & 1 \\
3UTRBERT & \cite{yang2023utrbert} & rna & Transformer encoder (BERT-style) & 86M-98M & 3-6-mer & 512 & 4 \\
UTR-LM & \cite{chu2023utrlm} & rna & Transformer encoder (ESM- or BERT-style) & 1.21M & nucleotide & 1,022 & 2 \\
Xpresso & \cite{agarwal2020xpresso} & dna & 1D CNN (promoter regression) & 0.11M & nucleotide & 10,500 & 1 \\
\bottomrule
\end{longtable}
\endgroup

  \end{landscape}

  \section{Extended Data 2: Standardized-checkpoint conversion summaries}\label{secA2}

  This table summarizes the output-level conversion checks behind the standardized-checkpoint count in the manuscript-reported inventory.
  The manuscript validation files used to generate this table include tested output categories, element-wise absolute-difference quantiles, tolerances and validation status.

  \begin{landscape}
    
\begingroup
\tiny
\setlength{\tabcolsep}{0.7pt}
\emergencystretch=2em
\begin{longtable}{@{}p{0.10\linewidth}p{0.21\linewidth}p{0.13\linewidth}p{0.055\linewidth}p{0.07\linewidth}p{0.07\linewidth}p{0.07\linewidth}p{0.07\linewidth}p{0.07\linewidth}p{0.07\linewidth}@{}}
\caption{\textbf{Conversion summaries for standardized model checkpoints.} Each row is one source-exposed output tensor tested for a standardized checkpoint counted in the manuscript-reported inventory. The columns report the distribution of element-wise absolute differences between the standardized and source-reference outputs: minimum, 25th percentile (Q25), median (Q50), 75th percentile (Q75), maximum and mean. Each checkpoint is evaluated on a deterministic reference input that exercises the model vocabulary. These columns are descriptive difference summaries rather than standalone pass/fail criteria; the manuscript validation files used to generate this table additionally store the absolute and relative tolerances, validation status and source-reference fixture or converted-output comparison metadata.}\label{tab:ed-checkpoints}\\
\toprule
Family & Checkpoint & Output & N & Min & Q25 & Q50 & Q75 & Max & Mean \\
\midrule
\endfirsthead
\toprule
Family & Checkpoint & Output & N & Min & Q25 & Q50 & Q75 & Max & Mean \\
\midrule
\endhead
a2zchromatin & a2zchromatin-accessibility & logits & 1 & 5.96e-08 & 5.96e-08 & 5.96e-08 & 5.96e-08 & 5.96e-08 & 5.96e-08 \\
a2zchromatin & a2zchromatin-methylation & logits & 1 & 1.49e-07 & 1.49e-07 & 1.49e-07 & 1.49e-07 & 1.49e-07 & 1.49e-07 \\
ablang & ablang-heavy & logits & 1200 & 0 & 0 & 0 & 1.79e-07 & 5.72e-06 & 2.82e-07 \\
ablang & ablang-light & logits & 1200 & 0 & 0 & 0 & 1.26e-07 & 6.68e-06 & 2.22e-07 \\
ablang2 & ablang2 & hidden\_states & 156000 & 0 & 7.15e-07 & 3.81e-06 & 1.53e-05 & 0.000389 & 1.38e-05 \\
ablang2 & ablang2 & logits & 650 & 0 & 1.01e-06 & 2.15e-06 & 3.87e-06 & 1.24e-05 & 2.79e-06 \\
amplify & amplify-120m & hidden\_states & 1996800 & 0 & 3.13e-07 & 9.54e-07 & 2.62e-06 & 0.000641 & 2.63e-06 \\
amplify & amplify-120m & logits & 3510 & 0 & 4.77e-07 & 1.91e-06 & 3.81e-06 & 1.53e-05 & 2.29e-06 \\
amplify & amplify-350m & hidden\_states & 3993600 & 0 & 4.77e-07 & 1.43e-06 & 3.81e-06 & 0.00252 & 4.32e-06 \\
amplify & amplify-350m & logits & 3510 & 0 & 4.77e-07 & 9.54e-07 & 1.91e-06 & 9.54e-06 & 1.41e-06 \\
aparent & aparent & logits & 1 & 2.38e-06 & 2.38e-06 & 2.38e-06 & 2.38e-06 & 2.38e-06 & 2.38e-06 \\
aparent2 & aparent2 & logits & 206 & 0 & 0 & 4.77e-07 & 9.54e-07 & 7.63e-06 & 1.04e-06 \\
basenji & basenji & coverage & 896 & 0 & 3.73e-09 & 1.49e-08 & 2.61e-08 & 2.53e-07 & 1.93e-08 \\
basset & basset & logits & 164 & 0 & 7.15e-07 & 1.91e-06 & 2.62e-06 & 4.05e-06 & 1.85e-06 \\
borzoi & borzoi-human & coverage & 6144 & 0 & 9.31e-10 & 2.33e-09 & 5.12e-09 & 7.75e-06 & 8.13e-09 \\
borzoi & borzoi-mouse & coverage & 6144 & 0 & 6.98e-10 & 1.75e-09 & 3.96e-09 & 2.74e-06 & 7.13e-09 \\
bpfold & bpfold & logits & 2500 & 0 & 1.12e-08 & 2.98e-08 & 5.96e-08 & 3.1e-06 & 6.83e-08 \\
bpnet & bpnet & profile\_logits & 8000 & 0 & 0 & 0 & 0 & 1.91e-06 & 1.44e-07 \\
bpnet & bpnet & count\_logits & 8 & 0 & 1.79e-07 & 3.58e-07 & 4.77e-07 & 9.54e-07 & 3.58e-07 \\
calm & calm & hidden\_states & 658944 & 0 & 1.04e-07 & 2.98e-07 & 6.56e-07 & 1.29e-05 & 4.84e-07 \\
calm & calm & logits & 2112 & 0 & 2.98e-07 & 7.15e-07 & 1.25e-06 & 1.14e-05 & 9.49e-07 \\
carp & carp-38m & hidden\_states & 2097152 & 0 & 0 & 0 & 0 & 0 & 0 \\
carp & carp-38m & logits & 3840 & 0 & 5.96e-08 & 1.19e-07 & 9.54e-07 & 1.34e-05 & 9.43e-07 \\
carp & carp-600k & hidden\_states & 262144 & 0 & 0 & 0 & 0 & 0 & 0 \\
carp & carp-600k & logits & 3840 & 0 & 0 & 0 & 0 & 5.72e-06 & 7.77e-08 \\
carp & carp-640m & hidden\_states & 9175040 & 0 & 0 & 0 & 0 & 0 & 0 \\
carp & carp-640m & logits & 3840 & 0 & 5.96e-08 & 1.19e-07 & 9.54e-07 & 3.62e-05 & 1.3e-06 \\
carp & carp-76m & hidden\_states & 4194304 & 0 & 0 & 0 & 0 & 0 & 0 \\
carp & carp-76m & logits & 3840 & 0 & 2.98e-08 & 1.19e-07 & 4.77e-07 & 1.53e-05 & 7.05e-07 \\
chrombpnet & chrombpnet & profile\_logits & 1000 & 0 & 1.34e-05 & 2.67e-05 & 4.24e-05 & 0.00013 & 3.04e-05 \\
chrombpnet & chrombpnet & count\_logits & 1 & 9.54e-07 & 9.54e-07 & 9.54e-07 & 9.54e-07 & 9.54e-07 & 9.54e-07 \\
deepcpgdna & deepcpgdna-hou2016-hcc & logits & 25 & 8.94e-08 & 5.96e-07 & 7.6e-07 & 1.01e-06 & 1.25e-06 & 7.7e-07 \\
deepcpgdna & deepcpgdna-hou2016-hepg2 & logits & 6 & 3.58e-07 & 3.58e-07 & 3.87e-07 & 5.51e-07 & 7.75e-07 & 4.77e-07 \\
deepcpgdna & deepcpgdna-hou2016-mesc & logits & 6 & 4.77e-07 & 4.77e-07 & 5.96e-07 & 7.15e-07 & 1.43e-06 & 7.15e-07 \\
deepcpgdna & deepcpgdna-smallwood2014-2i & logits & 12 & 0 & 2.24e-08 & 5.96e-08 & 9.69e-08 & 1.19e-07 & 5.96e-08 \\
deepcpgdna & deepcpgdna-smallwood2014-serum & logits & 18 & 0 & 7.45e-08 & 1.19e-07 & 2.24e-07 & 3.58e-07 & 1.52e-07 \\
deepmel & deepmel & logits & 24 & 0 & 1.79e-07 & 4.77e-07 & 4.77e-07 & 1.91e-06 & 4.17e-07 \\
deepsea & deepsea & logits & 919 & 2.38e-07 & 8.06e-05 & 0.000139 & 0.000192 & 0.000563 & 0.000169 \\
deepstarr & deepstarr & logits & 2 & 0 & 0 & 0 & 0 & 0 & 0 \\
deltasplice & deltasplice & probabilities & 1200 & 0 & 0 & 0 & 0 & 0 & 0 \\
deltasplice & deltasplice-human & probabilities & 1200 & 0 & 0 & 0 & 0 & 0 & 0 \\
dnabert & dnabert-3mer & hidden\_states & 638976 & 0 & 0 & 0 & 0 & 0 & 0 \\
dnabert & dnabert-3mer & logits & 4416 & 0 & 0 & 0 & 0 & 4.29e-06 & 8.45e-08 \\
dnabert & dnabert-3mer & vocab\_embeddings & 52992 & 0 & 0 & 0 & 0 & 0 & 0 \\
dnabert & dnabert-4mer & hidden\_states & 638976 & 0 & 0 & 0 & 0 & 0 & 0 \\
dnabert & dnabert-4mer & logits & 16704 & 0 & 0 & 0 & 0 & 0 & 0 \\
dnabert & dnabert-4mer & vocab\_embeddings & 200448 & 0 & 0 & 0 & 0 & 0 & 0 \\
dnabert & dnabert-5mer & hidden\_states & 638976 & 0 & 0 & 0 & 0 & 0 & 0 \\
dnabert & dnabert-5mer & logits & 65856 & 0 & 0 & 0 & 0 & 0 & 0 \\
dnabert & dnabert-5mer & vocab\_embeddings & 790272 & 0 & 0 & 0 & 0 & 0 & 0 \\
dnabert & dnabert-6mer & hidden\_states & 638976 & 0 & 0 & 0 & 0 & 0 & 0 \\
dnabert & dnabert-6mer & logits & 262464 & 0 & 0 & 0 & 0 & 0 & 0 \\
dnabert & dnabert-6mer & vocab\_embeddings & 3149568 & 0 & 0 & 0 & 0 & 0 & 0 \\
dnabert2 & dnabert2 & hidden\_states & 73728 & 0 & 0 & 0 & 0 & 0 & 0 \\
dnabert2 & dnabert2 & logits & 32768 & 0 & 0 & 0 & 0 & 0 & 0 \\
dnaberts & dnaberts & hidden\_states & 110592 & 0 & 0 & 0 & 0 & 0 & 0 \\
dnaberts & dnaberts & last\_hidden\_state & 9216 & 0 & 0 & 0 & 0 & 0 & 0 \\
dnaberts & dnaberts & pooler\_output & 768 & 0 & 7.45e-09 & 1.49e-08 & 2.24e-08 & 8.94e-08 & 1.58e-08 \\
enformer & enformer & coverage & 896 & 0 & 1.49e-08 & 3.17e-08 & 5.68e-08 & 7.93e-07 & 4.77e-08 \\
ernierna & ernierna & hidden\_states & 2665728 & 0 & 2.98e-08 & 8.94e-08 & 1.71e-07 & 2.21e-06 & 1.14e-07 \\
ernierna & ernierna & logits & 6141 & 0 & 1.19e-07 & 1.91e-06 & 2.86e-06 & 3.91e-05 & 1.84e-06 \\
ernierna & ernierna-ss & contact\_map & 70225 & 0 & 0 & 0 & 0 & 2.97e-05 & 1.46e-09 \\
framepool & framepool & logits & 1 & 4.77e-07 & 4.77e-07 & 4.77e-07 & 4.77e-07 & 4.77e-07 & 4.77e-07 \\
hal & hal & pooler\_output & 1 & 9.31e-10 & 9.31e-10 & 9.31e-10 & 9.31e-10 & 9.31e-10 & 9.31e-10 \\
hyenadna & hyenadna-large & hidden\_states & 2624000 & 0 & 2.98e-08 & 1.19e-07 & 3.58e-07 & 9.48e-06 & 2.59e-07 \\
hyenadna & hyenadna-large & logits & 11275 & 0 & 1.19e-07 & 9.54e-07 & 1.91e-06 & 3.43e-05 & 1.62e-06 \\
hyenadna & hyenadna-medium & hidden\_states & 1574400 & 0 & 7.45e-09 & 5.96e-08 & 1.79e-07 & 4.05e-06 & 1.22e-07 \\
hyenadna & hyenadna-medium & logits & 11275 & 0 & 0 & 4.77e-07 & 1.19e-06 & 1.24e-05 & 8.75e-07 \\
hyenadna & hyenadna-small & hidden\_states & 1049600 & 0 & 0 & 5.96e-08 & 1.34e-07 & 2.86e-06 & 9.21e-08 \\
hyenadna & hyenadna-small & logits & 11275 & 0 & 0 & 2.38e-07 & 9.54e-07 & 5.72e-06 & 6.92e-07 \\
hyenadna & hyenadna-tiny & hidden\_states & 524800 & 0 & 0 & 4.47e-08 & 1.19e-07 & 4.05e-06 & 9.44e-08 \\
hyenadna & hyenadna-tiny & logits & 11275 & 0 & 0 & 3.58e-07 & 9.54e-07 & 9.54e-06 & 6.93e-07 \\
malinois & malinois & logits & 3 & 0 & 0 & 0 & 5.96e-08 & 1.19e-07 & 3.97e-08 \\
maxentscan & maxentscan-score3 & logits & 1 & 0 & 0 & 0 & 0 & 0 & 0 \\
maxentscan & maxentscan-score5 & logits & 1 & 0 & 0 & 0 & 0 & 0 & 0 \\
mmsplice & mmsplice & logits & 1 & 0 & 0 & 0 & 0 & 0 & 0 \\
mpradragonn & mpradragonn & logits & 12 & 5.96e-08 & 8.94e-08 & 1.19e-07 & 2.53e-07 & 3.87e-07 & 1.7e-07 \\
mtsplice & mtsplice & logits & 56 & 0 & 7.45e-09 & 1.58e-08 & 2.98e-08 & 1.19e-07 & 1.93e-08 \\
openspliceai & openspliceai-arabidopsis.10000 & logits & 1200 & 0 & 0 & 2.38e-07 & 4.77e-07 & 2.86e-06 & 3.54e-07 \\
openspliceai & openspliceai-arabidopsis.2000 & logits & 1200 & 0 & 0 & 2.38e-07 & 4.77e-07 & 1.91e-06 & 2.94e-07 \\
openspliceai & openspliceai-arabidopsis.400 & logits & 1200 & 0 & 0 & 2.38e-07 & 4.77e-07 & 1.43e-06 & 3e-07 \\
openspliceai & openspliceai-arabidopsis.80 & logits & 1200 & 0 & 0 & 0 & 4.77e-07 & 1.91e-06 & 2.27e-07 \\
openspliceai & openspliceai-honeybee.10000 & logits & 1200 & 0 & 0 & 2.38e-07 & 4.77e-07 & 3.81e-06 & 3.59e-07 \\
openspliceai & openspliceai-honeybee.2000 & logits & 1200 & 0 & 0 & 2.98e-08 & 4.77e-07 & 1.91e-06 & 2.86e-07 \\
openspliceai & openspliceai-honeybee.400 & logits & 1200 & 0 & 0 & 0 & 4.77e-07 & 1.91e-06 & 2.26e-07 \\
openspliceai & openspliceai-honeybee.80 & logits & 1200 & 0 & 0 & 0 & 4.77e-07 & 1.91e-06 & 2.4e-07 \\
openspliceai & openspliceai-mane.10000 & logits & 1200 & 0 & 0 & 0 & 9.54e-07 & 3.81e-06 & 4.14e-07 \\
openspliceai & openspliceai-mane.2000 & logits & 1200 & 0 & 0 & 0 & 9.54e-07 & 3.81e-06 & 3.62e-07 \\
openspliceai & openspliceai-mane.400 & logits & 1200 & 0 & 0 & 4.77e-07 & 9.54e-07 & 3.81e-06 & 5.57e-07 \\
openspliceai & openspliceai-mane.80 & logits & 1200 & 0 & 0 & 4.77e-07 & 9.54e-07 & 3.81e-06 & 4.9e-07 \\
openspliceai & openspliceai-mouse.10000 & logits & 1200 & 0 & 0 & 0 & 9.54e-07 & 3.81e-06 & 4.43e-07 \\
openspliceai & openspliceai-mouse.2000 & logits & 1200 & 0 & 0 & 0 & 9.54e-07 & 3.81e-06 & 5.2e-07 \\
openspliceai & openspliceai-mouse.400 & logits & 1200 & 0 & 0 & 4.77e-07 & 9.54e-07 & 3.81e-06 & 4.94e-07 \\
openspliceai & openspliceai-mouse.80 & logits & 1200 & 0 & 0 & 0 & 4.77e-07 & 1.91e-06 & 2.98e-07 \\
openspliceai & openspliceai-zebrafish.10000 & logits & 1200 & 0 & 0 & 4.77e-07 & 4.77e-07 & 2.86e-06 & 3.03e-07 \\
openspliceai & openspliceai-zebrafish.2000 & logits & 1200 & 0 & 0 & 2.38e-07 & 4.77e-07 & 2.86e-06 & 2.92e-07 \\
openspliceai & openspliceai-zebrafish.400 & logits & 1200 & 0 & 0 & 0 & 4.77e-07 & 1.91e-06 & 2.47e-07 \\
openspliceai & openspliceai-zebrafish.80 & logits & 1200 & 0 & 0 & 2.38e-07 & 4.77e-07 & 2.86e-06 & 2.87e-07 \\
optimus5prime & optimus5prime & logits & 1 & 1.19e-07 & 1.19e-07 & 1.19e-07 & 1.19e-07 & 1.19e-07 & 1.19e-07 \\
optmrl & optmrl & logits & 1 & 0 & 0 & 0 & 0 & 0 & 0 \\
pangolin & pangolin & probabilities & 4800 & 0 & 0 & 0 & 0 & 0 & 0 \\
procapnet & procapnet & profile\_logits & 2000 & 0 & 0 & 0 & 0 & 0 & 0 \\
procapnet & procapnet & count\_logits & 1 & 0 & 0 & 0 & 0 & 0 & 0 \\
progen2 & progen2-base & hidden\_states & 1075200 & 0 & 3.91e-08 & 1.38e-07 & 3.58e-07 & 5.34e-05 & 2.73e-07 \\
progen2 & progen2-base & logits & 700 & 0 & 1.91e-06 & 3.81e-06 & 7.63e-06 & 1.53e-05 & 4.25e-06 \\
progen2 & progen2-bfd90 & hidden\_states & 2112000 & 0 & 5.96e-08 & 2.16e-07 & 4.92e-07 & 0.000122 & 3.61e-07 \\
progen2 & progen2-bfd90 & logits & 700 & 0 & 1.19e-07 & 2.98e-07 & 7.15e-07 & 5.72e-06 & 5.77e-07 \\
progen2 & progen2-large & hidden\_states & 2112000 & 0 & 5.96e-08 & 1.9e-07 & 4.62e-07 & 0.000198 & 3.17e-07 \\
progen2 & progen2-large & logits & 700 & 0 & 5.96e-08 & 2.09e-07 & 4.77e-07 & 2.48e-05 & 6.39e-07 \\
progen2 & progen2-medium & hidden\_states & 1075200 & 0 & 5.96e-08 & 1.79e-07 & 4.47e-07 & 7.34e-05 & 3.33e-07 \\
progen2 & progen2-medium & logits & 700 & 0 & 0 & 7.63e-06 & 1.53e-05 & 4.58e-05 & 9.44e-06 \\
progen2 & progen2-oas & hidden\_states & 1075200 & 0 & 4.47e-08 & 1.49e-07 & 3.58e-07 & 2.19e-05 & 2.41e-07 \\
progen2 & progen2-oas & logits & 700 & 0 & 0 & 7.63e-06 & 1.53e-05 & 8.01e-05 & 1.12e-05 \\
progen2 & progen2-small & hidden\_states & 332800 & 0 & 7.45e-09 & 6.71e-08 & 2.09e-07 & 4.58e-05 & 1.66e-07 \\
progen2 & progen2-small & logits & 700 & 0 & 0 & 3.81e-06 & 7.63e-06 & 3.05e-05 & 5.11e-06 \\
progen2 & progen2-xlarge & hidden\_states & 3379200 & 0 & 1.19e-07 & 3.58e-07 & 7.9e-07 & 0.000183 & 6.08e-07 \\
progen2 & progen2-xlarge & logits & 700 & 0 & 0 & 3.05e-05 & 8.39e-05 & 0.000679 & 7.41e-05 \\
proteinbert & proteinbert & hidden\_states & 99840 & 0 & 1.86e-08 & 4.47e-08 & 8.94e-08 & 1.25e-06 & 6.87e-08 \\
proteinbert & proteinbert & logits & 3380 & 0 & 4.77e-07 & 1.19e-06 & 2.15e-06 & 2.86e-05 & 2.21e-06 \\
proteinbert & proteinbert & annotation\_logits & 8943 & 0 & 1.91e-06 & 4.77e-06 & 8.58e-06 & 3.43e-05 & 5.81e-06 \\
ribonanzanet & ribonanzanet & logits\_2a3 & 50 & 1.19e-07 & 9.72e-07 & 1.83e-06 & 3.78e-06 & 1.34e-05 & 2.83e-06 \\
ribonanzanet & ribonanzanet & logits\_dms & 50 & 8.94e-08 & 1.14e-06 & 2.77e-06 & 4.85e-06 & 1.68e-05 & 3.77e-06 \\
ribonanzanet & ribonanzanet-deg & logits\_reactivity & 50 & 5.96e-08 & 1.36e-06 & 3.06e-06 & 5.17e-06 & 1.25e-05 & 3.54e-06 \\
ribonanzanet & ribonanzanet-deg & logits\_deg\_Mg\_pH10 & 50 & 2.98e-07 & 2.27e-06 & 3.78e-06 & 5.81e-06 & 3.04e-05 & 5.63e-06 \\
ribonanzanet & ribonanzanet-deg & logits\_deg\_pH10 & 50 & 2.38e-07 & 2.53e-06 & 4.26e-06 & 7.13e-06 & 2.73e-05 & 5.23e-06 \\
ribonanzanet & ribonanzanet-deg & logits\_deg\_Mg\_50C & 50 & 2.09e-07 & 1.92e-06 & 3.24e-06 & 5.09e-06 & 1.76e-05 & 3.98e-06 \\
ribonanzanet & ribonanzanet-deg & logits\_deg\_50C & 50 & 5.96e-08 & 2.47e-06 & 4.14e-06 & 6.89e-06 & 1.93e-05 & 4.88e-06 \\
ribonanzanet & ribonanzanet-drop & logits\_2a3 & 1 & 0.00446 & 0.00446 & 0.00446 & 0.00446 & 0.00446 & 0.00446 \\
ribonanzanet & ribonanzanet-drop & logits\_dms & 1 & 0.00189 & 0.00189 & 0.00189 & 0.00189 & 0.00189 & 0.00189 \\
ribonanzanet & ribonanzanet-ss & logits\_2a3 & 50 & 5.96e-08 & 5.85e-07 & 1.15e-06 & 1.41e-06 & 6.47e-06 & 1.29e-06 \\
ribonanzanet & ribonanzanet-ss & logits\_dms & 50 & 0 & 6.26e-07 & 1.13e-06 & 2.02e-06 & 4.71e-06 & 1.41e-06 \\
ribonanzanet & ribonanzanet-ss & logits\_ss & 2500 & 0 & 5.91e-05 & 0.000109 & 0.000164 & 0.000599 & 0.000117 \\
rinalmo & rinalmo-giga & logits & 1144 & 0 & 0 & 0 & 0 & 7.63e-06 & 2.32e-07 \\
rinalmo & rinalmo-giga-ss & logits & 2500 & 0 & 0 & 0 & 0 & 0 & 0 \\
rinalmo & rinalmo-mega & logits & 1144 & 0 & 0 & 9.54e-07 & 1.91e-06 & 1.14e-05 & 1.2e-06 \\
rinalmo & rinalmo-micro & logits & 1144 & 0 & 9.31e-08 & 9.54e-07 & 1.91e-06 & 6.68e-06 & 1.39e-06 \\
rnabert & rnabert & hidden\_states & 36000 & 0 & 0 & 2.33e-09 & 2.98e-08 & 1.79e-06 & 2.62e-08 \\
rnabert & rnabert & logits\_lm & 300 & 1.67e-10 & 3.39e-09 & 7.3e-09 & 1.06e-08 & 2.18e-08 & 7.71e-09 \\
rnabert & rnabert & logits\_sa & 2 & 0 & 0 & 0 & 0 & 0 & 0 \\
rnabert & rnabert & logits\_ss & 400 & 1.05e-11 & 2.43e-09 & 6.94e-09 & 1.02e-08 & 2.3e-08 & 6.87e-09 \\
rnaernie & rnaernie & logits & 520 & 0 & 3.24e-07 & 1.88e-06 & 2.86e-06 & 3.81e-05 & 2.8e-06 \\
rnafm & rnafm & hidden\_states & 432640 & 0 & 4.77e-07 & 1.67e-06 & 4.29e-06 & 0.000122 & 3.14e-06 \\
rnafm & rnafm & logits & 1144 & 0 & 9.54e-07 & 2.21e-06 & 4.77e-06 & 4.39e-05 & 3.24e-06 \\
rnafm & rnafm-mrna & hidden\_states & 299520 & 0 & 4.77e-07 & 1.91e-06 & 4.29e-06 & 0.000305 & 3.22e-06 \\
rnafm & rnafm-mrna & logits & 1260 & 0 & 9.54e-07 & 2.15e-06 & 4.17e-06 & 5.53e-05 & 3.77e-06 \\
rnafm & rnafm-ss & hidden\_states & 432640 & 0 & 4.77e-07 & 1.67e-06 & 4.29e-06 & 0.000122 & 3.14e-06 \\
rnafm & rnafm-ss & logits & 1144 & 0 & 9.54e-07 & 2.21e-06 & 4.77e-06 & 4.39e-05 & 3.24e-06 \\
rnamsm & rnamsm & hidden\_states & 430848 & 0 & 0 & 0 & 0 & 0 & 0 \\
rnamsm & rnamsm & logits & 612 & 0 & 0 & 0 & 2.98e-07 & 1.91e-06 & 3.04e-07 \\
scbasset & scbasset & logits & 2034 & 0 & 9.54e-07 & 2.38e-06 & 4.29e-06 & 1.24e-05 & 2.78e-06 \\
splicebert & splicebert & hidden\_states & 32256 & 0 & 1.49e-08 & 5.96e-08 & 1.34e-07 & 1.43e-06 & 9.17e-08 \\
splicebert & splicebert & logits & 90 & 0 & 4.75e-08 & 2.91e-07 & 1.91e-06 & 1.53e-05 & 1.87e-06 \\
splicebert & splicebert & vocab\_embeddings & 5120 & 0 & 0 & 0 & 0 & 0 & 0 \\
splicebert & splicebert-human.510 & hidden\_states & 32256 & 0 & 0 & 4.47e-08 & 1.42e-07 & 1.19e-06 & 9.32e-08 \\
splicebert & splicebert-human.510 & logits & 90 & 0 & 9.69e-08 & 5.96e-07 & 1.91e-06 & 1.38e-05 & 1.19e-06 \\
splicebert & splicebert-human.510 & vocab\_embeddings & 5120 & 0 & 0 & 0 & 0 & 0 & 0 \\
splicebert & splicebert.510 & hidden\_states & 32256 & 0 & 1.49e-08 & 7.45e-08 & 1.79e-07 & 1.19e-06 & 1.13e-07 \\
splicebert & splicebert.510 & logits & 90 & 0 & 0 & 7.15e-07 & 2.26e-06 & 1.34e-05 & 2.21e-06 \\
splicebert & splicebert.510 & vocab\_embeddings & 5120 & 0 & 0 & 0 & 0 & 0 & 0 \\
spotrna & spotrna & contact\_map & 2500 & 0 & 0 & 0 & 1.86e-09 & 2.71e-06 & 1.44e-08 \\
sptransformer & sptransformer & logits & 7200 & 0 & 0 & 0 & 0 & 0 & 0 \\
ufold & ufold & logits & 2500 & 0 & 2.86e-06 & 5.72e-06 & 1.14e-05 & 6.48e-05 & 8.21e-06 \\
utrbert & utrbert-3mer & hidden\_states & 638976 & 0 & 0 & 0 & 0 & 0 & 0 \\
utrbert & utrbert-3mer & logits & 4416 & 0 & 0 & 0 & 0 & 3.81e-06 & 4.88e-08 \\
utrbert & utrbert-3mer & vocab\_embeddings & 52992 & 0 & 0 & 0 & 0 & 0 & 0 \\
utrbert & utrbert-4mer & hidden\_states & 638976 & 0 & 0 & 0 & 0 & 0 & 0 \\
utrbert & utrbert-4mer & logits & 16704 & 0 & 0 & 0 & 0 & 0 & 0 \\
utrbert & utrbert-4mer & vocab\_embeddings & 200448 & 0 & 0 & 0 & 0 & 0 & 0 \\
utrbert & utrbert-5mer & hidden\_states & 638976 & 0 & 0 & 0 & 0 & 0 & 0 \\
utrbert & utrbert-5mer & logits & 65856 & 0 & 0 & 0 & 0 & 0 & 0 \\
utrbert & utrbert-5mer & vocab\_embeddings & 790272 & 0 & 0 & 0 & 0 & 0 & 0 \\
utrbert & utrbert-6mer & hidden\_states & 638976 & 0 & 0 & 0 & 0 & 0 & 0 \\
utrbert & utrbert-6mer & logits & 262464 & 0 & 0 & 0 & 0 & 0 & 0 \\
utrbert & utrbert-6mer & vocab\_embeddings & 3149568 & 0 & 0 & 0 & 0 & 0 & 0 \\
utrlm & utrlm-mrl & logits & 468 & 1.26e-05 & 4.1e-05 & 8.34e-05 & 0.000163 & 0.000406 & 0.000108 \\
utrlm & utrlm-te\_el & logits & 468 & 0 & 5.36e-07 & 1.91e-06 & 3.81e-06 & 7.15e-05 & 3.6e-06 \\
xpresso & xpresso & logits & 1 & 1.19e-07 & 1.19e-07 & 1.19e-07 & 1.19e-07 & 1.19e-07 & 1.19e-07 \\
\bottomrule
\end{longtable}
\endgroup

  \end{landscape}

  \section{Extended Data 3: Curated dataset resources and public repository inventory}\label{secA3}

  This table enumerates the 39 public dataset repositories through which the 16 curated dataset resources are released, grouping each repository under its source resource and listing the variant that distinguishes it from sibling repositories (sequence-length caps, alignment construction, species or dataset-defined splits) together with its molecule, task geometry and label type.
  It documents processed dataset assets and does not imply that every arbitrary cross-dataset comparison is leakage-free.

  \begin{landscape}
    
\begingroup
\tiny
\setlength{\tabcolsep}{1pt}
\emergencystretch=2em
\begin{longtable}{@{}p{0.12\linewidth}p{0.05\linewidth}p{0.28\linewidth}p{0.22\linewidth}p{0.04\linewidth}p{0.10\linewidth}@{}}
\caption{\textbf{Curated dataset resources and their public repository inventory.} The 16 curated dataset resources are released through 39 public dataset repositories on the MultiMolecule Hugging Face Hub organization. Each row is one public repository, grouped under its source resource, with the variant that distinguishes it from sibling repositories of the same resource: sequence-length caps ($\le$N nt), alignment construction, species or dataset-defined splits. Molecule: R = RNA, D = DNA. Task: S = sequence-level, T = token-level, C = contact-level; Bin = binary, MC = multiclass, Reg = regression. The molecule and task columns report the source resource's geometry, which the variant slices. The Ref column cites each resource's source publication. The table documents processed dataset assets rather than asserting that every cross-dataset comparison is leakage-free.}\label{tab:ed-datasets}\\
\toprule
Resource & Ref & Public repository & Variant & Mol. & Task \\
\midrule
\endfirsthead
\toprule
Resource & Ref & Public repository & Variant & Mol. & Task \\
\midrule
\endhead
ArchiveII & \cite{samanbooy2022archiveii} & multimolecule/archiveii & full set & R & C $\times$ Bin \\
ArchiveII & \cite{samanbooy2022archiveii} & multimolecule/archiveii.512 & $\le$512 nt & R & C $\times$ Bin \\
ArchiveII & \cite{samanbooy2022archiveii} & multimolecule/archiveii.1024 & $\le$1024 nt & R & C $\times$ Bin \\
bpRNA-1m & \cite{danaee2018bprna} & multimolecule/bprna & full set & R & C $\times$ Bin \\
bpRNA-1m & \cite{danaee2018bprna} & multimolecule/bprna-90 & $<$90\% similarity subset & R & C $\times$ Bin \\
bpRNA-new & \cite{sato2021mxfold2} & multimolecule/bprna-new & full set & R & C $\times$ Bin \\
bpRNA-spot & \cite{singh2019spotrna} & multimolecule/bprna-spot & full set & R & C $\times$ Bin \\
bpRNA-spot & \cite{singh2019spotrna} & multimolecule/bprna-spot-0 & bpRNA-1m split (TR0, VL0 and TS0) & R & C $\times$ Bin \\
bpRNA-spot & \cite{singh2019spotrna} & multimolecule/bprna-spot-1 & PDB transfer-learning split & R & C $\times$ Bin \\
bpRNA-spot & \cite{singh2019spotrna} & multimolecule/bprna-spot-2 & NMR-only eval split (TS2) & R & C $\times$ Bin \\
CASP-RNA & \cite{kryshtafovych2023casp} & multimolecule/casp-rna & full set & R & C $\times$ Bin \\
CHANRG & \cite{chen2026chanrg} & multimolecule/chanrg & full set & R & S $\times$ MC \\
EternaBench-CM & \cite{waymentsteele2022eternabench} & multimolecule/eternabench-cm & full set & R & T $\times$ Reg \\
EternaBench-External & \cite{waymentsteele2022eternabench} & multimolecule/eternabench-external.300 & $\le$300 nt & R & T $\times$ Reg \\
EternaBench-External & \cite{waymentsteele2022eternabench} & multimolecule/eternabench-external.600 & $\le$600 nt & R & T $\times$ Reg \\
EternaBench-External & \cite{waymentsteele2022eternabench} & multimolecule/eternabench-external.900 & $\le$900 nt & R & T $\times$ Reg \\
EternaBench-External & \cite{waymentsteele2022eternabench} & multimolecule/eternabench-external.1200 & $\le$1200 nt & R & T $\times$ Reg \\
EternaBench-Switch & \cite{waymentsteele2022eternabench} & multimolecule/eternabench-switch & full set & R & T $\times$ Reg \\
GENCODE & \cite{harrow2006gencode} & multimolecule/gencode-human & human & R, D & T $\times$ Bin \\
GENCODE & \cite{harrow2006gencode} & multimolecule/gencode-mouse & mouse & R, D & T $\times$ Bin \\
IPknot++ & \cite{sato2022ipknot} & multimolecule/ipknot\_plus\_plus & full set, no alignment & R & C $\times$ Bin \\
IPknot++ & \cite{sato2022ipknot} & multimolecule/ipknot\_plus\_plus-ref & Rfam reference alignment & R & C $\times$ Bin \\
IPknot++ & \cite{sato2022ipknot} & multimolecule/ipknot\_plus\_plus-mafft & MAFFT alignment & R & C $\times$ Bin \\
Rfam & \cite{griffithsjones2003rfam} & multimolecule/rfam & full set & R & S $\times$ MC \\
RIVAS & \cite{rivas2012rivas} & multimolecule/rivas & full set & R & C $\times$ Bin \\
RIVAS & \cite{rivas2012rivas} & multimolecule/rivas-a & Set A & R & C $\times$ Bin \\
RIVAS & \cite{rivas2012rivas} & multimolecule/rivas-b & Set B (Rfam) & R & C $\times$ Bin \\
RNAcentral & \cite{bateman2011rnacentral} & multimolecule/rnacentral & full set & R & T $\times$ Bin, C $\times$ Bin \\
RNAcentral & \cite{bateman2011rnacentral} & multimolecule/rnacentral-modifications & modification-annotated subset & R & T $\times$ Bin, C $\times$ Bin \\
RNAcentral & \cite{bateman2011rnacentral} & multimolecule/rnacentral.512 & $\le$512 nt & R & T $\times$ Bin, C $\times$ Bin \\
RNAcentral & \cite{bateman2011rnacentral} & multimolecule/rnacentral.1024 & $\le$1024 nt & R & T $\times$ Bin, C $\times$ Bin \\
RNAcentral & \cite{bateman2011rnacentral} & multimolecule/rnacentral.2048 & $\le$2048 nt & R & T $\times$ Bin, C $\times$ Bin \\
RNAcentral & \cite{bateman2011rnacentral} & multimolecule/rnacentral.4096 & $\le$4096 nt & R & T $\times$ Bin, C $\times$ Bin \\
RNAcentral & \cite{bateman2011rnacentral} & multimolecule/rnacentral.8192 & $\le$8192 nt & R & T $\times$ Bin, C $\times$ Bin \\
RNAStrAlign & \cite{tan2017turbofold} & multimolecule/rnastralign & full set & R & C $\times$ Bin \\
RNAStrAlign & \cite{tan2017turbofold} & multimolecule/rnastralign.512 & $\le$512 nt & R & C $\times$ Bin \\
RNAStrAlign & \cite{tan2017turbofold} & multimolecule/rnastralign.1024 & $\le$1024 nt & R & C $\times$ Bin \\
RYOS & \cite{waymentsteele2022ryos} & multimolecule/ryos-1 & round 1 (len 107) & R & T $\times$ Reg \\
RYOS & \cite{waymentsteele2022ryos} & multimolecule/ryos-2 & round 2 (len 130) & R & T $\times$ Reg \\
\bottomrule
\end{longtable}
\endgroup

  \end{landscape}

  \section{Extended Data 4: Prediction-pipeline registry}\label{secA4}

  This table lists the registered task interface for each user-facing prediction pipeline counted in the manuscript-reported inventory.

  \begin{landscape}
    
\begingroup
\tiny
\setlength{\tabcolsep}{1pt}
\emergencystretch=2em
\begin{longtable}{@{}p{0.18\linewidth}p{0.27\linewidth}p{0.27\linewidth}p{0.28\linewidth}@{}}
\caption{\textbf{Prediction-pipeline registry.} Each row lists a user-facing prediction interface: the accepted input form, returned output fields and postprocessing behavior of its registered task interface. Each pipeline binds a configurable default standardized checkpoint and accepts any compatible checkpoint; the default may change across releases and is resolved from the registry rather than fixed here. Every registered pipeline is covered by a unit test of registration, default-model metadata, AutoModel-class resolution and postprocessing behavior. The ten registered pipelines are deployed through nine public Hugging Face Spaces; the regulatory-track and regulatory-profile pipelines share one regulatory-signal demo surface. Hosted smoke tests are treated as implementation checks rather than as display-level results.}\label{tab:ed-pipelines}\\
\toprule
Pipeline & Accepted input & Returned output & Postprocessing \\
\midrule
\endfirsthead
\toprule
Pipeline & Accepted input & Returned output & Postprocessing \\
\midrule
\endhead
rna-secondary-structure & RNA sequence string or batch of sequences & sequence, secondary\_structure, optional contact\_map & contact logits or probabilities to contact map and dot-bracket notation \\
splice-site & DNA or RNA sequence string & sequence, splice-site scores, optional top-k positions & position-level channel scores with threshold and top-k filtering \\
splice-variant-effect & reference and alternative sequence strings & reference\_sequence, alternative\_sequence, delta scores, optional top-k effects & alternative-minus-reference splice effect scores \\
polyadenylation & DNA sequence string or mapping with sequence & sequence, score, cleavage\_distribution or channel scores & polyadenylation scores and cleavage-distribution formatting \\
regulatory-activity & DNA sequence string or mapping with optional numeric features & sequence, channel scores & whole-sequence regulatory activity scores \\
regulatory-track & DNA sequence string & sequence, genomic bins, channel scores & binned regulatory track scores \\
regulatory-profile & DNA sequence string & sequence, base-resolution profile scores & per-base regulatory profile reconstruction \\
regulatory-variant-effect & reference and alternative DNA sequence strings, optional feature vectors & reference\_sequence, alternative\_sequence, delta scores, optional top-k effects & alternative-minus-reference regulatory effect scores \\
methylation & DNA sequence string & sequence, methylation channel scores & whole-sequence methylation prediction scores \\
mean-ribosome-load & 5-prime UTR sequence string & sequence, mean\_ribosome\_load score & mean-ribosome-load scalar formatting \\
\bottomrule
\end{longtable}
\endgroup

  \end{landscape}

  \FloatBarrier

\end{appendices}

\bibliography{sn-bibliography}

\end{document}